\documentclass[twocolumn]{aastex631}
\usepackage{amsmath}

\shorttitle{Repeating TDEs}
\shortauthors{Liu et al.}

\newcommand{\md}{\mathrm d}
\newcommand{\dmde}{\mathrm dM/\mathrm dE}

\newcommand{\mbh}{M_{\mathrm{BH}}}
\newcommand{\rT}{r_{\mathrm{T}}}
\newcommand{\rp}{r_{\mathrm{p}}}
\newcommand{\tpeak}{t_{\mathrm{peak}}}
\newcommand{\tdyn}{t_{\mathrm{dyn}}}

\newcommand{\mosfit}{\texttt{MOSFiT}}
\usepackage[normalem]{ulem}
\usepackage{CJK}
\graphicspath{{./}{figures/}}
\begin{document}
\title{Tidal disruption events from eccentric orbits and lessons learned from the noteworthy ASASSN-14ko}

\correspondingauthor{Chang Liu}
\email{ptg.cliu@u.northwestern.edu}

\author[0000-0002-7866-4531]{\begin{CJK*}{UTF8}{gbsn}Chang Liu (刘畅)\end{CJK*}}
\affiliation{Kavli Institute for Astronomy and Astrophysics, Peking University, 5 Yiheyuan Road, Beijing 100871, China}
\affil{Department of Physics and Astronomy, Northwestern University, 2145 Sheridan Rd, Evanston, IL 60208, USA}
\affil{Center for Interdisciplinary Exploration and Research in Astrophysics (CIERA), Northwestern University, 1800 Sherman Ave, Evanston, IL 60201, USA}

\author[0000-0001-6350-8168]{Brenna Mockler}
\affiliation{Department of Astronomy and Astrophysics, University of California, Santa Cruz, CA 95064, USA}

\author[0000-0003-2558-3102]{Enrico Ramirez-Ruiz}
\affiliation{Department of Astronomy and Astrophysics, University of California, Santa Cruz, CA 95064, USA}

\author[0000-0003-0381-1039]{Ricardo Yarza}
\affiliation{Department of Astronomy and Astrophysics, University of California, Santa Cruz, CA 95064, USA}

\author[0000-0001-8825-4790]{Jamie A.P. Law-Smith}
\affiliation{Center for Astrophysics $|$ Harvard \& Smithsonian, Cambridge, MA 02138, USA}

\author[0000-0002-9802-9279]{Smadar Naoz}
\affiliation{Department of Physics and Astronomy, University of California, Los Angeles, CA 90095, USA}
\affiliation{Mani L. Bhaumik Institute for Theoretical Physics, Department of Physics and Astronomy, UCLA, Los Angeles, CA 90095, USA}

\author[0000-0002-7854-1953]{Denyz Melchor}
\affiliation{Department of Physics and Astronomy, University of California, Los Angeles, CA 90095, USA}
\affiliation{Mani L. Bhaumik Institute for Theoretical Physics, Department of Physics and Astronomy, UCLA, Los Angeles, CA 90095, USA}

\author[0000-0003-0984-4456]{Sanaea Rose}
\affiliation{Department of Physics and Astronomy, University of California, Los Angeles, CA 90095, USA}
\affiliation{Mani L. Bhaumik Institute for Theoretical Physics, Department of Physics and Astronomy, UCLA, Los Angeles, CA 90095, USA}

\begin{abstract}
    Stars grazing supermassive black holes (SMBHs) on bound orbits may survive tidal disruption, causing periodic flares. Inspired by the recent discovery of the periodic nuclear transient ASASSN-14ko, a promising candidate for a repeating tidal disruption event (TDE), we study the tidal deformation of stars approaching SMBHs on eccentric orbits. With both analytical and hydrodynamics methods, we show the overall tidal deformation of a star is similar to that in a parabolic orbit provided that the eccentricity is above a critical value. This allows one to make use of existing simulation libraries from parabolic encounters to calculate the mass fallback rate in eccentric TDEs. We find the flare structures of eccentric TDEs show a complicated dependence on both the SMBH mass and the orbital period. For stars orbiting SMBHs with relatively short periods, we predict significantly shorter-lived duration flares than those in parabolic TDEs, which can be used to predict repeating events if the mass of the SMBH can be independently measured. Using an adiabatic mass loss model, we study the flare evolution over multiple passages, and show the evolved stars can survive many more passages than main sequence stars. We apply this theoretical framework to the repeating TDE candidate ASASSN-14ko and suggest that its recurrent flares originate from a moderately massive ($M\gtrsim 1\,\mathrm{M_\odot}$), extended ({likely $\sim$10\,$\mathrm{R_\odot}$}), evolved star on a grazing, bound orbit around the SMBH. Future hydrodynamics simulations of multiple tidal interactions will enable realistic models on the individual flare structure and the evolution over multiple flares.
\end{abstract}

\section{Introduction}
Most massive black holes, such as Sgr A* at the center of the Milky Way, are quiescent, accreting
gas at a very low rate \citep{Ho_nuclear_2008}. Although its mass is $4\times 10^6 \,\mathrm{M_\odot}$ \citep{Ghez_1998, Genzel_2000},
Sgr A* has a luminosity of only about $10^2\,L_\odot$ \citep{Baganoff_SgrA_2003}, less than many individual giant stars. Yet,
massive black holes inhabit galactic center environments of extreme stellar density \citep{Boker_2008}. In  these galactic nuclei, stars trace dangerously wandering
orbits dictated by the combined gravitational force of a central, supermassive black hole (SMBH) and all of the surrounding stars \citep{2005PhR...419...65A}. When the occasional star plunges too close to the SMBH and is disrupted by its tidal field \citep{hills_possible_1975}, it can fuel an extremely luminous accretion flare \citep{rees_tidal_1988}. Stars can be tidally disrupted at any evolutionary stage and can include main sequence stars and evolved stars \citep{macleod_tidal_2012, Arcavi_2014,Law-Smith_WD_2017}. The accompanying flare is not only a definitive sign of the
presence of an otherwise quiescent SMBH but also a powerful diagnostic of the properties of the disrupted star \citep{Ramirez-Ruiz_2009,Lodato_2009,guillochon_hydrodynamical_2013} and particularly if the composition of the debris can be inferred \citep{Kochanek_abundance_2016, gallegos-garcia_tidal_2018,2022ApJ...924...70M}. 

Observational surveys are now routinely  discovering the optical, X-ray and radio counterparts of these events at rates of tens per year \citep{Komossa_2015, Auchettl_2017, van_Velzen_2021}. The observed rates of tidal disruption events (TDEs) hold important discriminatory power over both the dynamical mechanisms operating in galactic nuclei and the nature of their underlying stellar populations \citep{French_2020, 2022ApJ...924...70M}. However, the dynamical mechanisms that feed stars into disruptive orbits remain highly uncertain.
There have been several theories in addition to the standard two-body relaxation  \citep{1976MNRAS.176..633F,1999MNRAS.309..447M,Stone_rates_2016} put forward to explain the rates of TDEs, from an overdensity of stars near the SMBH \citep{Stone_rates_2016,2017ApJ...850...22L,Graur_2018}, to star formation in eccentric disks around SMBHs \citep{Madigan_2018, Wernke_2019}, to the three-body interaction between stellar binaries and SMBHs \citep{Hills_1988,Sari_2010,Cufari_2022}, to the presence of SMBH binaries \citep{Ivanov_2005, Chen_2009, Chen_2011, Stone_2011, Li_2015}.  
What these theories share in common is that the disrupted stars come from within the radius of influence of the SMBH. Deep in the potential of the SMBH,
these stars gravitationally interact with one another coherently, in contrast to two-body
relaxation, resulting in a much more rapid angular momentum evolution \citep{Rauch_resonant_1996, Kocsis_2015}. 
In these scenarios, disrupted   
orbits are more likely to  be bound and highly eccentric, which could give rise to periodic outbursts if the star can survive  \citep{macleod_spoon-feeding_2013,Nixon_2022}. These expectations  have become more relevant in light of the recently claimed repeated-flaring TDE ASASSN-14ko \citep{Payne_ASASSN_2021}.  This recent detection and its multi-flare characterization motivates our study of stellar disruption in eccentric orbits.

In this paper we explore via a hydrodynamical study  how mass fallback rates are expected to vary   between the widely studied  parabolic case and the eccentric scenario in Section~\ref{sec:Sim}, while Section~\ref{sec:Analyt} sets the stage by introducing the problem analytically. In Section \ref{Sec:LC} we compute the properties of eccentric TDE light curves from our simulations, and use them to derive a relationship between the time of peak of the TDE flare and the SMBH mass. In Section \ref{sec:PhaseSpace}, we use this relationship to predict whether a particular TDE flare is likely to repeat. While our hydrodynamical study is restricted to individual passages, in Section~\ref{Sec:MF} we present a model aimed at constraining the stellar structure of stars experiencing multiple tidal encounters. Finally our constraints on the nature of the star fueling   ASASSN-14ko are presented in Section~\ref{Sec:dis} along with  a discussion of future prospects and lessons learned. 

\section{Setting the Stage}
\label{sec:Analyt}
When a disrupted star is initially in a parabolic orbit, as assumed for canonical TDEs, roughly half of the stellar debris is bound to the SMBH. However, repeating TDEs are guaranteed  to occur in eccentric orbits when the surviving  remnant and all  of the disrupted material is bound to the SMBH. This happens when the orbital eccentricity of the star is below the critical value \citep{hayasaki_tidal_2012}
\begin{equation}
    e_\mathrm{cri}=1-2\frac{q^{-1/3}}{\beta},
    \label{eq:hay}
\end{equation}
where $q\equiv \mbh/M_*$ is the mass ratio, and $\beta\equiv \rT/\rp$ is the impact parameter of the encounter, the ratio of the tidal radius $\rT\equiv (\mbh/M_*)^{1/3}R_*$ and the pericenter distance $\rp$. In the case of a complete disruption with $e< e_\mathrm{cri}$, all of the shredded material is bound to the SMBH and is expected to end up being fully accreted. The resultant TDE is predicted to drastically increase in luminosity when the bulk of debris returns to the pericenter, which takes place within approximately an orbital period for the center of mass of the disrupted debris \citep{Dai_2013, hayasaki_classification_2018, Park_2020}. For partial disruptions with $e< e_\mathrm{cri}$, the majority of the star may survive multiple pericenter passages and the decay rate after each disruption depends crucially on the hydrodynamical evolution of the debris stream, which is influenced by both the surviving core and the SMBH. As the ratio between the tidal and self-gravitational forces  evolves persistently over the encounter, the relationship between the  distribution of binding energy across the star and $\beta$ is more intricate \citep{guillochon_hydrodynamical_2013} than the one used to derive Equation~(\ref{eq:hay}). This means that simulations are needed and  they must cover far more than a few dynamical timescales after the disruption, with  the final functional form the mass return rate not being established
until the star is many hundreds of tidal radii away from the SMBH \citep{guillochon_consequences_2011,Coughlin_2019, Miles_2020}.

\begin{figure}
    \centering
    \includegraphics[width=\linewidth]{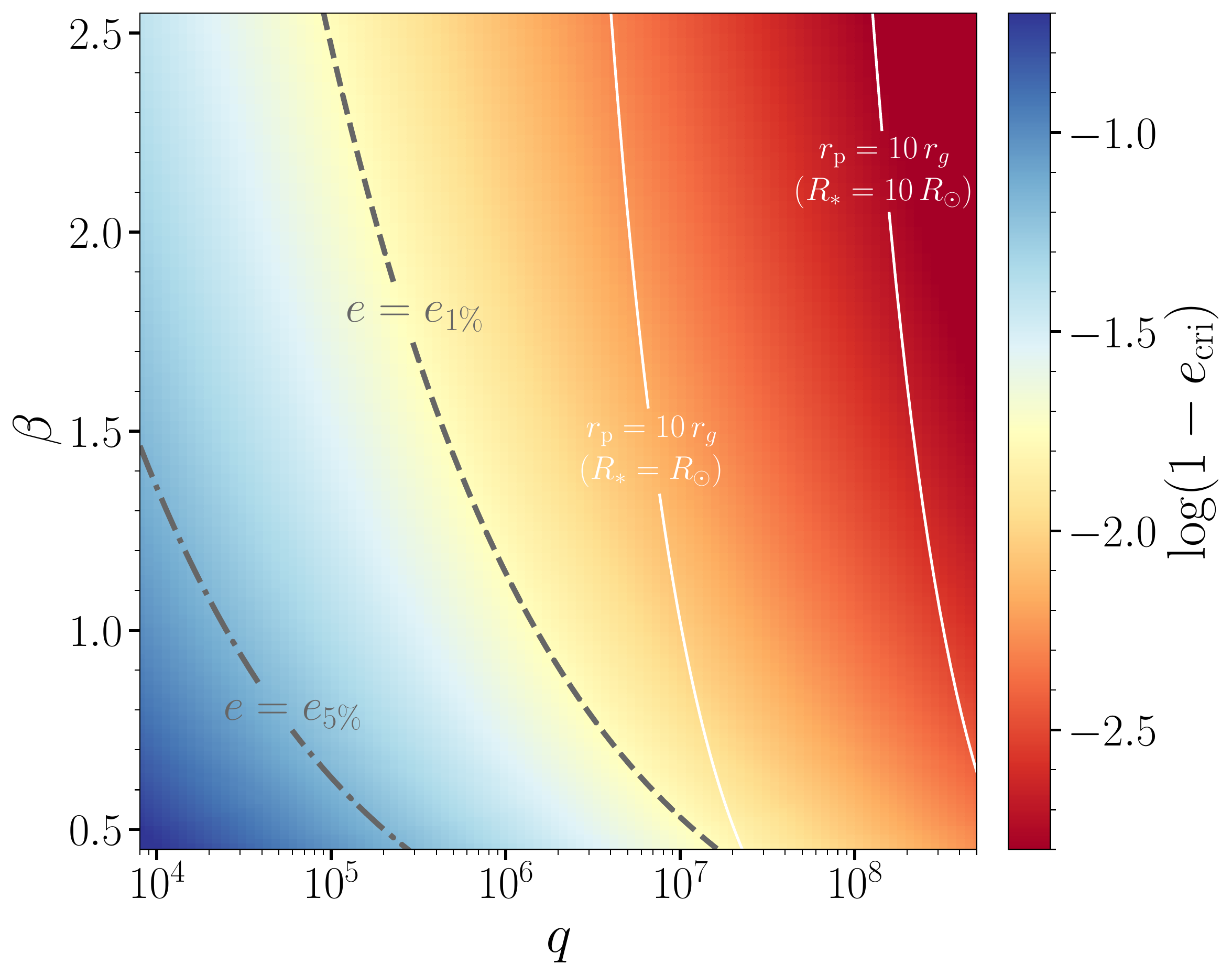}
    \caption{Critical eccentricity $e_{\mathrm{cri}}$ from Equation~(\ref{eq:hay}), at which all the debris is exactly bound to the SMBH, as a function of $\beta$ and $q$. The dashed and dash-dotted curves correspond to the eccentricities $e=e_{1\%}$ and $e=e_{5\%}$, respectively. To the right side to the dashed curve, $e_{1\%}<e_\mathrm{cri}$ always holds. The same is true for $e_{5\%}$ to the right of the dash-dotted line. For stars in elliptical orbits with $e<e_\mathrm{cri}$, only those in the region to the right of the dash-dotted curve ($e_{1\%}<e<e_\mathrm{cri}$) or dashed curve ($e_{5\%}<e<e_\mathrm{cri}$) are similarly distorted to those in parabolic orbits. {The two solid curves mark the critical $\beta$ given $q$ at which the pericenter distance $\rp$ of a 1\,$\mathrm{R_\odot}$ (10\,$\mathrm{R_\odot}$) star is 10\,$\mathrm{r_\mathrm{g}}$. When the star is disrupted by a more massive SMBH, the relativistic effects cannot be neglected.}}
    \label{fig:analy_cri_ecc}
\end{figure}

The disrupted material closest to the SMBH, i.e., the inner tidal tail, returns to pericenter ahead of the surviving core in a partial disruption. {When $e<e_{\rm cri}$, the outer tidal tail is also bound to the SMBH. Part of the tail remains within the surviving core's Hill sphere, such that the material will be re-accreted by the star within a few dynamical timescales. The remaining tidal tail material  will eventually return to the pericenter, but since it is much less tightly bound to the SMBH compared to the inner tail (which causes the major flare), it will return long after the surviving core does. As a result, the mass fallback rate is reduced, leading to a lower luminosity, and  the time interval between {two consecutive encounters between the stellar remnant and the SMBH} is dictated by the orbital period of the star rather than the fallback time, which is significantly longer. After multiple flares, the material from the outer tails will likely overlap and interact with each other. This  could generate a continuous level of low mass accretion between flares \citep{macleod_spoon-feeding_2013}. As such, the fallback of material from the outer tails  will probably  generate a low-luminosity background level of radiation with some mild  degree of variability. This is in sharp contrast to the fallback of material arising  from the inner tails, which is responsible for causing the bright individual flares. Modeling the background radiation requires simulations of multiple passages, which is beyond the scope of this paper. In what follows we thus assume that the mass fallback drastically terminates when the surviving core returns to pericenter, which is in contrast to the parabolic case in which the accretion rate does not terminate abruptly but continues to steadily decrease \citep{guillochon_hydrodynamical_2013}.} In a repeating TDE, as shown in Section~\ref{sec:Sim},  we expect a much steeper decay in the luminosity after each encounter given the distribution of binding energies of the material.

Still, when the eccentricity is close to unity, it is possible that the debris structure after an {\it eccentric} tidal disruption remains similar to those undergoing deformation in parabolic orbits, even if the resultant  light curves are much steeper and shorter-lived. To explore this scenario and estimate the degree of distortion after an eccentric or parabolic disruption, we assess the overall impact of tidal distortions with the time integration, $I$, of the tidal force, $F_\mathrm{T}(r)=4G\mbh R_*/r^3$, 
\begin{equation}
\label{eq:I}
\begin{split}
    I&=\int_{t(r_\mathrm{min})}^{t(r_\mathrm{max})}F_T(t,e)\md t\\
    &=4G\mbh R_*\int_{t(r_\mathrm{min})}^{t(r_\mathrm{max})}r^{-3}(t,e)\md t,
\end{split}
\end{equation}
as the star travels from pericenter to apocenter for eccentric TDEs, or to infinity for parabolic ones. The relative deviation in $I$ with respect to the parabolic case is solely a function of $e$ because Keplerian orbits are scale-free. For a sun-like star, a 1\% deviation in $I$ requires an eccentricity $e_{1\%}=0.9825$, while a 5\% deviation requires $e_\mathrm{ 5\%}=0.9317$. Sizable deviations to the tidal deformation from cases in parabolic orbits are only expected for eccentricities lower than $e_{5\%}$.

In Figure~\ref{fig:analy_cri_ecc}, we show how $e_\mathrm{cri}$ varies with the depth of the encounter and the mass ratio between the SMBH and the disrupted object assuming a sun-like star. The two curves correspond to the eccentricities $e=e_\mathrm{ 1\%}$ (dashed) and $e=e_\mathrm{ 5\%}$ (dash-dotted). For stars undergoing eccentric disruptions ($e<e_\mathrm{cri}$), only those in the region to the right of the dash-dotted curve ($e_{1\%}<e<e_\mathrm{cri}$) or dashed curve ($e_{5\%}<e<e_\mathrm{cri}$) are similarly distorted to those in parabolic orbits, even though the light curves can look dramatically different given the expected differences in the orbital energy distribution of the returning debris.

{We note that a non-relativistic approach is not  applicable when the pericenter distance $\rp$ is close to the event horizon of the SMBH. For this reason, in Figure~\ref{fig:analy_cri_ecc} we also plot the critical eccentricities given $\beta$, $q$, and the stellar radius $R_*$, at which $\rp=10\,r_\mathrm{g}$, where $r_\mathrm{g}\equiv G\mbh/c^2$ is the gravitational radius. When $\rp>10\,r_\mathrm{g}$, studies on relativistic encounters \citep{Cheng_2014,Servin_2017,Tejeda_2017,Gafton_2019,Stone_2019,Ryu_2020d} have shown that the difference between the mass fallback rates calculated in Newtonian and relativistic simulations is {$\lesssim$10--20\%}. For more extended stars (such as evolved stars), our analytical estimation remains applicable even for the more massive SMBHs.}

\section{Simulations}
\label{sec:Sim}

\subsection{The need for hydrodynamics simulations}
\label{subsec:Sim-Model}
The analytical considerations presented in Section~\ref{sec:Analyt} provide a useful framework  for interpreting the results of hydrodynamics simulations for sun-like stars disrupted by a SMBH in eccentric orbits. Hydrodynamics simulations are particularly important for deriving the fallback rate $\dot M(t)$, which depends sensitively  on the binding energy distribution $\md E/\md M$ of the stellar debris as \citep[e.g.,][]{rees_tidal_1988, Evans_TDE_1989, Phinney_1989},
\begin{equation}
\label{eq:mdot}
    \dot M=\frac{\md M}{\md E}\frac{\md E}{\md t}=\frac23\frac{\md M}{\md E}\left(\frac{\pi^2G^2\mbh^2}{2}\right)^{1/3}t^{-5/3}.
\end{equation}
The surviving core and the material bound to it will not contribute to the mass fallback \citep{Lodato_2009, Coughlin_2019,law-smith_tidal_2019,Ryu_2020c}, and are thus excluded from our calculations. Details on the binding energy distribution and the structure of the debris before and after disruption are discussed in Section~\ref{subsec:Sim-Result}. 

In this paper we closely follow the approach of \citet{law-smith_stellar_2020} in the Stellar TDEs with Abundances and Realistic Structures (STARS) library, which was used to  evaluate the mass fallback rate of main-sequence (MS) stars disrupted by SMBHs. The density profile and chemical abundances profile of a star are modeled using the 1D stellar evolution code \texttt{MESA} \citep{Paxton_mesa_2011}. We use a sun-like star model with zero age main sequence (ZAMS) mass of 1\,$\mathrm{M_\odot}$, and an age of 0.445\,Gyr, when the stellar radius is roughly $\mathrm{R_\odot}$. The profile is then mapped into the 3D adaptive-mesh refinement (AMR) hydrodynamics code \texttt{FLASH} \citep{fryxell_flash_2000}. The major difference with \citet{law-smith_stellar_2020} is that we place the CoM of the star in an elliptical orbit with a given eccentricity rather than placing the star in a parabolic orbit.

We study three impact parameters, $\beta=0.8, 1, 2$. The critical impact parameter of a full disruption $\beta_\mathrm{full}$ is $\sim$3 for a sun-like star in a parabolic orbit \citep{law-smith_stellar_2020,Ryu_2020c}.  Our simulations thus cover a large variety of partial disruptions and our results for the corresponding  parabolic encounters are completely consistent with \citet{law-smith_stellar_2020}.

For a given $\beta$ we test three simulation setups, whose orbits were intentionally selected to describe the variety of stellar disruptions expected around a SMBH with properties similar to the one that powered ASASSN-14ko, whose mass, $M_{\rm BH}$, is estimated to be around $7 \times 10^7\,\mathrm{M_\odot}$\footnote{We note here that our results are  applicable to any non-relativistic encounter since all disruption quantities can be scaled  with the BH mass \citep{guillochon_hydrodynamical_2013,mockler_weighing_2019}.} \citep{Payne_ASASSN_2021}. 
The three setups are: (i) encounters in parabolic orbits; (ii) encounters in eccentric orbits with a fixed orbital period of 114$\,$d, which is the orbital period estimated for ASASSN-14ko \citep{Payne_ASASSN_2021}; and  (iii) encounters in eccentric orbits with a fixed eccentricity $e=0.9$, well below $e_{5\%}$. For case (ii), all three orbits tested, each with a different $\beta$ for a sun-like star, have $e>e_{1\%}$ (for $\beta=0.8,1,2$, $e=0.9914, 0.9931, 0.9965$, respectively). We assume the orbit of the star to be Keplerian and do not allow the orbital period or eccentricity to change, since the period of ASASSN-14ko evolves rather slowly \citep[period derivative $\dot P=-0.0017\pm0.0003$;][]{Payne_ASASSN_2021}.

\subsection{Results}
\label{subsec:Sim-Result}
\begin{figure}
    \centering
    \includegraphics[width=\linewidth]{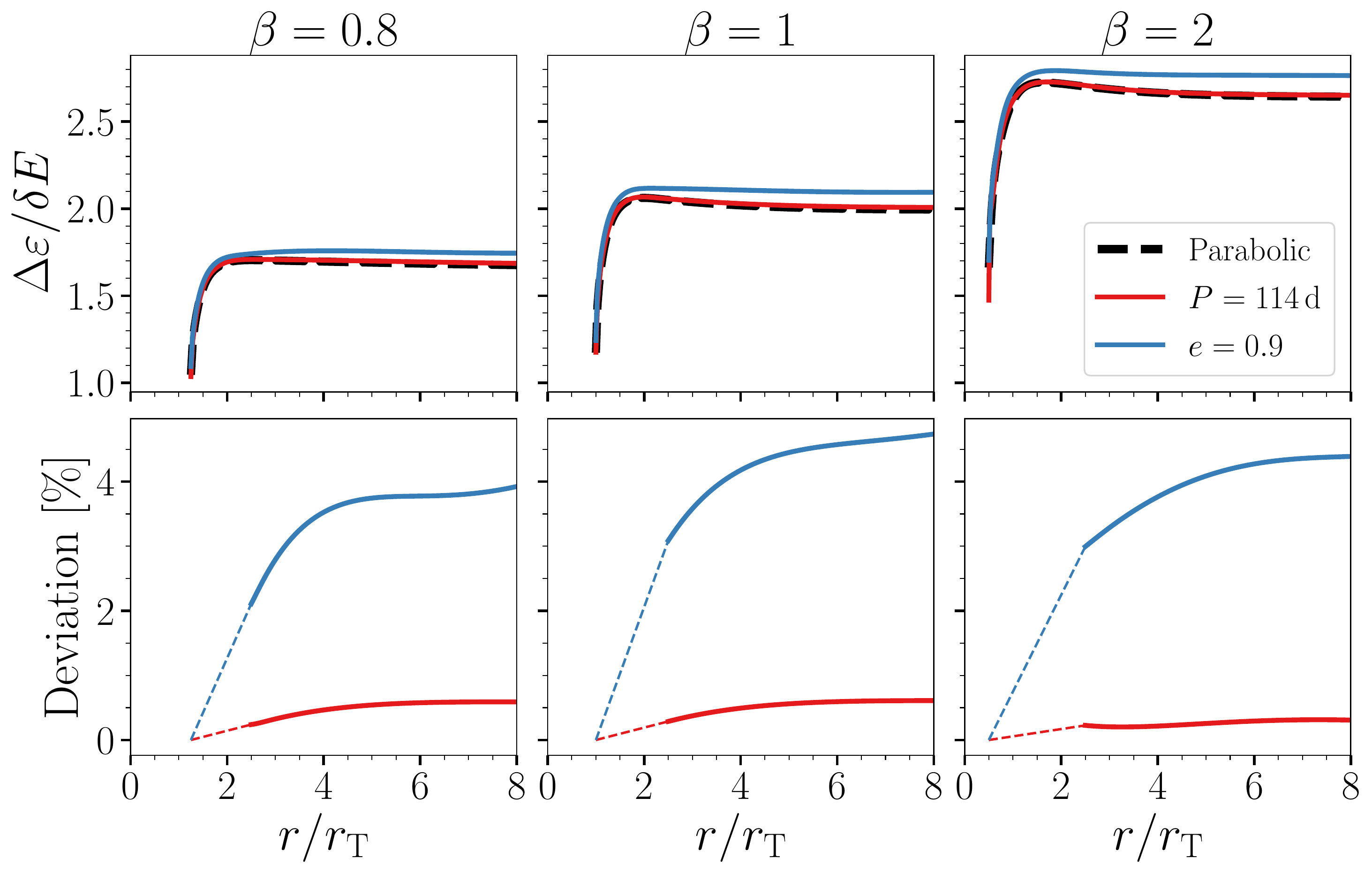}
    \caption{Energy dispersion as a function of radial distance for different $\beta$ and $e$. For each case, we calculate the difference between the largest and smallest specific binding energy, $\Delta \varepsilon$, across the region above one critical density such that $\sim$2/3 of the mass which is not in the self-bound core is encompassed. The energy range is normalized by the typical energy dispersion for a sun-like star. In the upper panels, curves of different colors correspond to the normalized energy dispersion after crossing the pericenter in (i) parabolic orbits (black); (ii) orbits with a period of 114$\,$d, or $e=0.9914, 0.9931, 0.9965$ for $\beta=0.8,1,2$ (red); and (iii) orbits with $e=0.9$ (blue).  In the lower panels the deviations from parabolic TDEs of the two eccentric cases is shown. The results are smoothed with B-splines. When the star has just passed the pericenter, the data are too sparse for a valid B-spline fitting due to the extremely large orbital velocity, so we simply assume the deviation is negligible at the pericenter, and grows almost linearly when $\rp<r<2.5\,\rp$. This linear growth is shown in dashed curves.}
    \label{fig:dEde}
\end{figure} 

With the goal of paving the way for the presentation of our numerical study, we employ the difference between the largest and smallest specific binding energy across the debris as a measure of the degree of tidal distortion, $\Delta\varepsilon$. In Figure~\ref{fig:dEde} we plot the evolution of $\Delta\varepsilon$ during various  tidal encounters. In all cases, $\Delta\varepsilon$ is normalized by the typical energy dispersion of a sun-like star at its tidal radius, $\delta E\equiv G\mbh R_*/\rT^2$. 

The difference in the energy dispersion calculated for cases (i) and (ii) is unnoticeable ($\lesssim$0.5\%), but case (iii) deviates markedly  from the  parabolic case, showing an energy dispersion that is larger by $\sim$4\%. It is worth noting that  in  simulations of the disruption of a star in an eccentric orbit, the  resulting tidal deformation is smaller than those predicted by the energy-freezing model, which was used to calculate Equation (\ref{eq:hay}) and (\ref{eq:I}). This simple formalism predicts a deviation  $\gtrsim$5\% in the binding energy dispersion for $e=0.9$  while we derive $\sim$4\%. The reader is refer to \citet{guillochon_hydrodynamical_2013} for a complementary  discussion of the limitations of the energy-freezing model for  parabolic encounters with varying $\beta$.

\begin{figure}
        \centering
        \includegraphics[width=\linewidth]{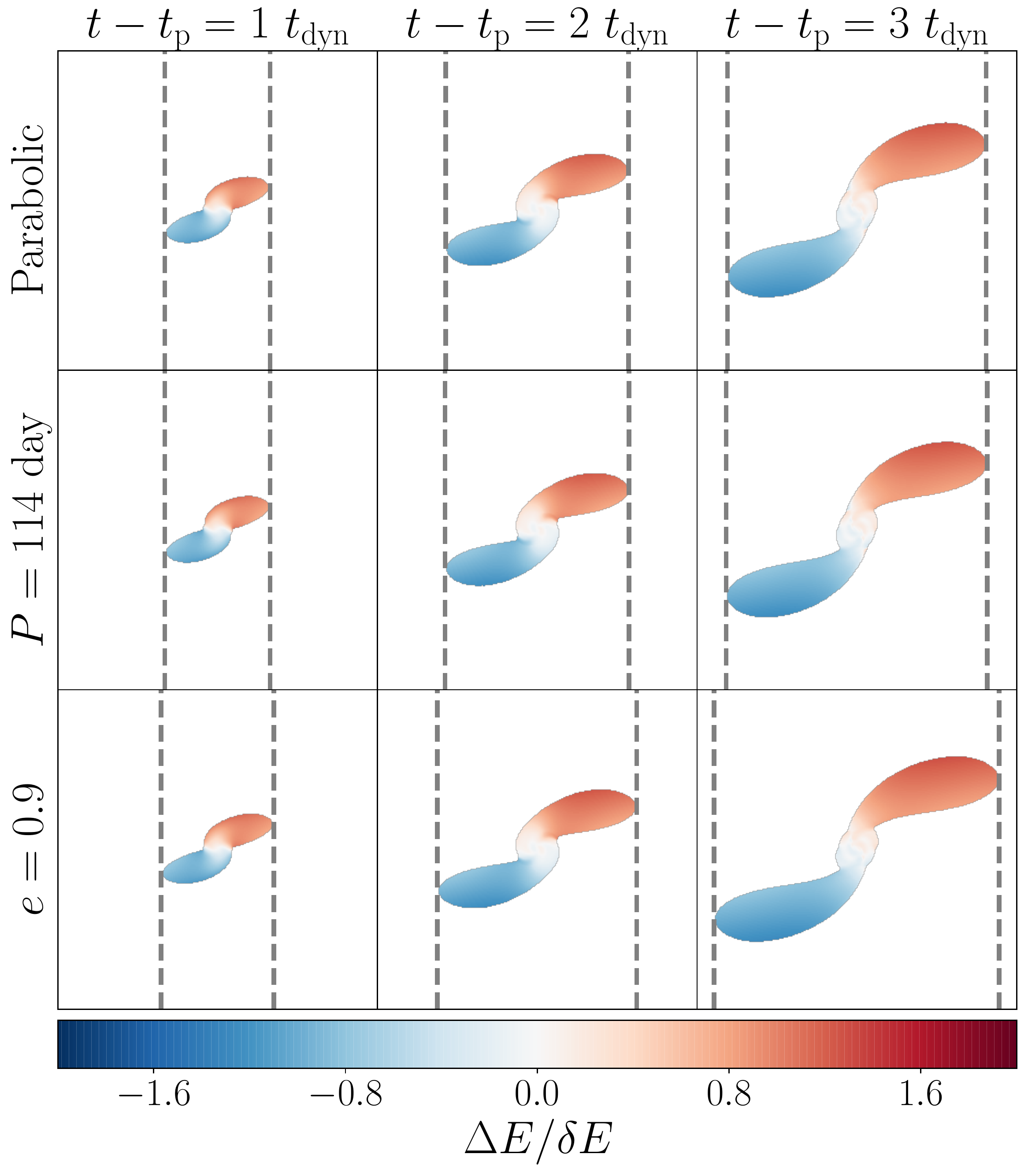}
        \caption{Dispersion in specific binding energy to the SMBH in the stellar debris for a fixed $\beta=1$. Shown are the 2D projections weighted by density. The three rows correspond to the cases in parabolic orbits (upper panels), orbits with a period of 114$\,$d (middle panels), and orbits with $e=0.9$ (lower panels). In each row, we show the snapshot for the debris 1, 2, and 3\,$t_\mathrm{dyn}$ after the encounter, respectively. For each snapshot, again we include $\sim$2/3 of the stellar mass that is not self-bound.  $\Delta E$ is the relative specific orbital energy with respect to the center of mass of the debris. The width of each snapshot is 12\,$\mathrm{R_\odot}$. Edges of the encompassed region are marked with vertical dashed lines.}
        \label{fig:Simu_Snapshot_Beta}
    \end{figure}
    \begin{figure}
        \centering
        \includegraphics[width=\linewidth]{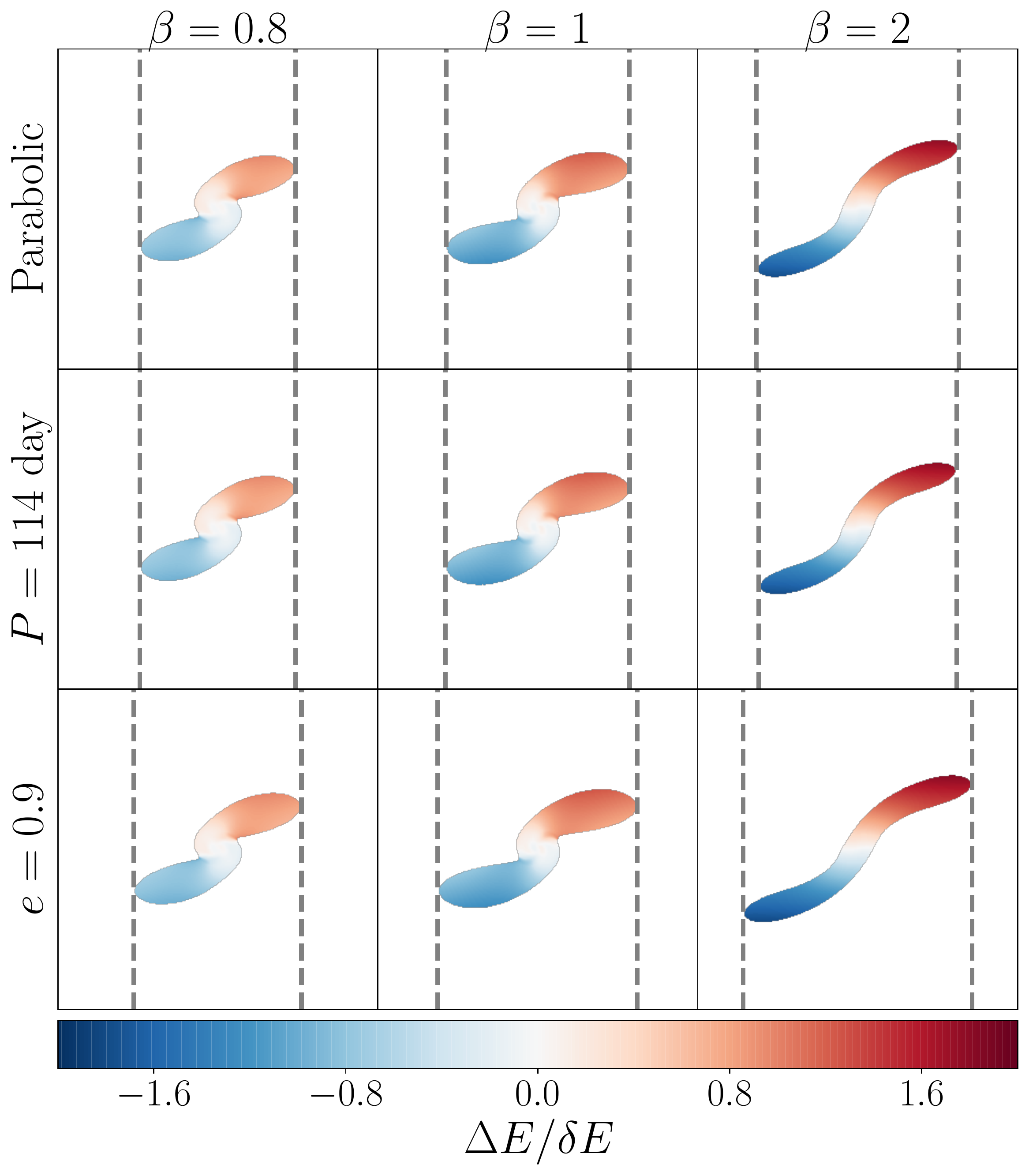}
        \caption{Dispersion in specific binding energy in the stellar debris $2\,t_\mathrm{dyn}$ after the encounter. Shown are the 2D projections weighted by density. The three rows again correspond to the cases in parabolic orbits (upper panels), orbits with a period of 114 d (middle panels), and orbits with $e=0.9$ (lower panels), below the critical eccentricity $e_\mathrm{ 5\%}$. For each column, $\beta=0.8$, 1, and 2, respectively. We use the same density threshold selection and same notations as that in Figure~\ref{fig:Simu_Snapshot_Beta}.}
        \label{fig:Simu_Snapshot_Time}
    \end{figure}

We plot the debris structure and energy distribution in Figures~\ref{fig:Simu_Snapshot_Beta} and \ref{fig:Simu_Snapshot_Time}. Shown are  the 2D snapshots of binding energy to the SMBH in the debris shortly after  the pericenter passage time, $t_\mathrm p$.  
As illustrated in Figure~\ref{fig:dEde},  the energy dispersion across the debris develops swiftly after the encounter, and converges in a few  dynamical timescales, $\tdyn\equiv(R_*^3/GM_*)^{1/2}$. So we only show snapshots  about $3\,\tdyn$ after $t_\mathrm p$.

In Figure~\ref{fig:Simu_Snapshot_Beta}, we show simulation snapshots at different times for a fixed $\beta$ but varying eccentricity. In Figure~\ref{fig:Simu_Snapshot_Time}, on the other hand, we show  simulation snapshots for varying $\beta$  taken at $2\,\tdyn$ after $t_\mathrm{p}$. To facilitate comparison, only the relative binding energy with respect to the CoM of the debris, $\Delta E$, is displayed, which is again normalized to the typical energy dispersion $\delta E$. In both figures, only the bottom panels with an eccentricity well below the critical value show a significantly deviation when compared to the parabolic encounter. 

\begin{figure*}
    \centering
    \includegraphics[width=\linewidth]{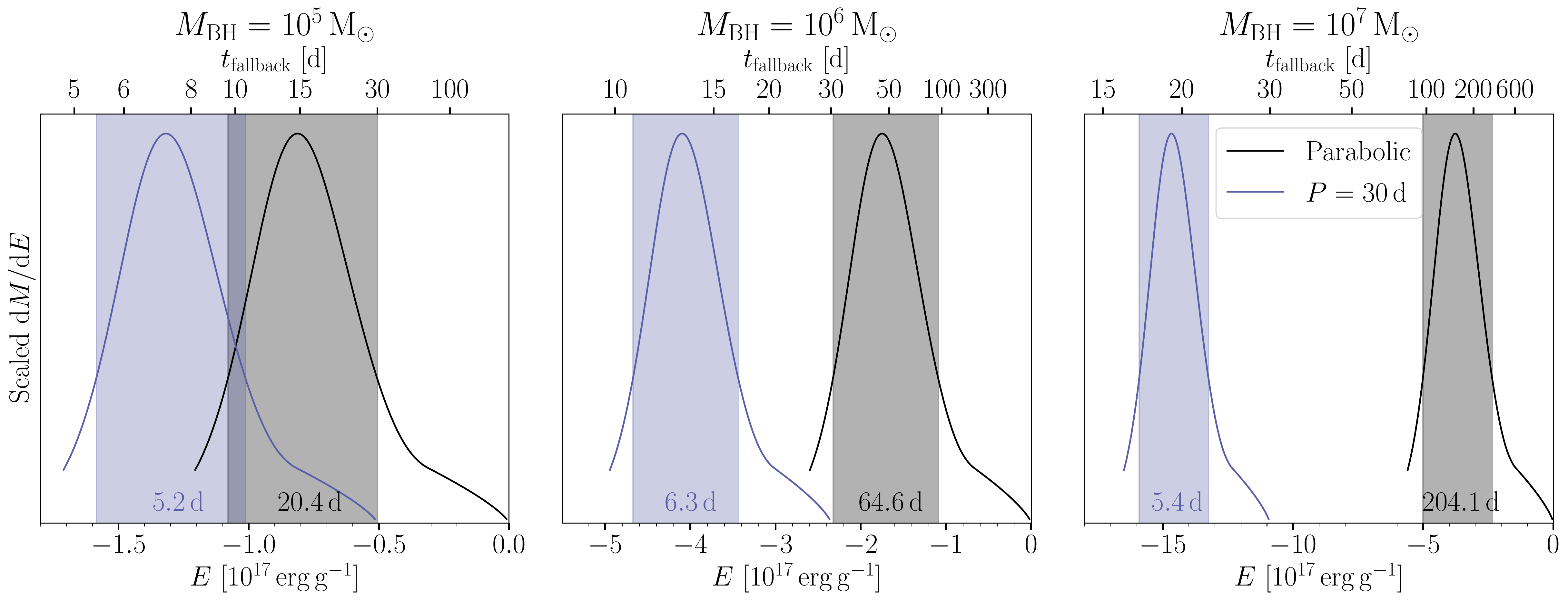}
    \caption{{The distribution of specific binding energies in eccentric TDEs (Equation~\ref{eq:dmde}). Here we show how shifting the specific binding energy changes the duration of the flare. We adopt the ${\md M}/{\md E}$ of the inner tidal tail from a sun-like star in a $\beta=1$ parabolic orbit around a SMBH of $10^6\,\mathrm{M_\odot}$ from the STARS library (black curve in the center panel). The least bound material in the tail has a specific binding energy of zero, thus the fallback time $t_\mathrm{fallback}$ is infinity. The blue curve has exactly the same shape as the black one, but the stripped star is on an eccentric, 30-day orbit, so the least bound material has a $t_\mathrm{fallback}=30\,\mathrm{d}$. The corresponding $t_\mathrm{fallback}$ for different $E$ are labeled in the axes on the top. In the left (right) panel, we re-scale the parabolic result for a $10^5\,\mathrm{M_\odot}$ ($10^7\,\mathrm{M_\odot}$) SMBH following Equation~(\ref{eq:scaling_relation}), then again shift it to match that of a 30-day orbit. The shaded regions show the $e$-folding full widths of ${\md M}/{\md E}$, where ${\md M}/{\md E}$ is greater than $1/e$ of its peak value. When shifted to eccentric orbits, ${\md M}/{\md E}$ retains its width in terms of $E$, but shows a much smaller width in terms of $t_\mathrm{fallback}$ (defined as $t_\mathrm{w}$), indicating a shorter duration of the mass fallback.}}
    \label{fig:dMde}
\end{figure*}

For an eccentricity similar to that derived for ASASSN-14ko, the disrupted star has a similarly shaped specific binding energy distribution when compared to the parabolic case, even though the mean value of the distribution has shifted significantly. In this case, when approximating the binding energy distribution with the simulation results for parabolic encounters, the actual energy spread will be underestimated by less than a few percent. This is no longer the case when $e$ approaches $e_{5\%}$. In these extreme cases, tailored  simulations for eccentric encounters are needed in order to derive $\md M/\md E$. Intriguingly, when $e>e_\mathrm{ 5\%}$, $\md M/\md E$  can be easily calculated from the parabolic library of stellar disruptions by making use of the fact that the entire distribution has been shifted to the star's center of mass (CoM),
\begin{equation}
    E_{\mathrm{CoM}}=-\left(\frac{\pi^2G^2\mbh^2}{2P^2}\right)^{1/3},
\end{equation}
where $P$ is the orbital period. The  energy dispersion for an eccentric disruption can then be derived by making use of the following relation 
\begin{equation}
\label{eq:dmde}
    \frac{\md M}{\md E}\left(E\right)=\left(\frac{\md M}{\md E}\right)_0\left(E-E_{\mathrm{CoM}}\right),
\end{equation}
where the subscript $0$ denotes for the original distribution in a parabolic orbit, or when $E_{\mathrm{CoM}}=0$. {Figure~\ref{fig:dMde} shows the shifting of the distribution of specific energies. We adopt the ${\md M}/{\md E}$ of the inner tidal tail stripped from a sun-like star in a $\beta=1$ parabolic orbit from the STARS library, and shift it to match that of an eccentric, 30-day orbit. Since the fallback time $t_\mathrm{fallback}\propto |E|^{-3/2}$ due to Kepler's third law, the tidal tail of an eccentric TDE will return to the pericenter much faster than that from a parabolic TDE.} This formalism allows for light curves to be calculated using a library of parabolic simulations like the one presented in \citet{law-smith_stellar_2020} for a broad range of stellar ages and masses.

We note that we only simulate the disruption of stars that have not been tidally perturbed, meaning that we are modeling the first pericenter passage since the unlucky star is captured. In more general cases, the mass fallback from a star surviving multiple encounters might not be well approximated by any models in existing simulations for parabolic TDEs, because the stellar structure might have been significantly deformed. The limitations of this individual encounter model and the need for hydrodynamics simulations of multiple passages are discussed in Section~\ref{subsec:multi_pass}.

\section{{Mass fallback curves} from eccentric TDE\lowercase{s}}\label{Sec:LC}

The bolometric luminosity in an individual flare depends sensitively on the mass fallback rate, and the dependence reduces to an approximate proportional relationship {if an accretion disk is formed promptly or if the emission is mainly liberated by shocks during the circularization processes such as stream collisions within the infalling debris  \citep{Piran_2015, Bonnerot_2016, Bonnerot_2020}. If the radiation is mainly accretion-powered, two conditions have to be met}: (i) the accretion rate follows the mass fallback rate, meaning that the tidally stripped material circularizes promptly after returning to the pericenter, and the viscous delay in the accretion disk is subdominant (see Section~\ref{sec:a14ko_dens} for more discussion on circularization and viscous delay); and (ii) the luminosity follows the accretion rate, and the radiation efficiency remains relatively constant during a flare. In optical TDEs,  multi-band light curves fit that make use of theoretical mass fallback curves provide a reasonable description of the total bolometric luminosity and deduce that  the  viscous delay is negligible in most, if not all cases \citep{mockler_weighing_2019, Nicholl_2022}. Motivated by this, in this section we discuss the overall shapes and the the typical timescales of the mass fallback curves arising from eccentric TDEs. These curves should provide a reasonable estimate of the bolometric luminosity of the expected flaring events irrespective of the dissipation process.

It is easy to scale the mass fallback rate $\dot M(t)$ for parabolic TDEs in the Newtonian regime. This is because the typical mass fallback rate and flare timescale both depend on the SMBH mass, the stellar mass, and the interior structure of the disrupted star \citep{Ramirez-Ruiz_2009,Lodato_2009, Haas_2012, guillochon_hydrodynamical_2013, law-smith_tidal_2019, Ryu_2020b, Ryu_2020c}, such that
\begin{equation}
\label{eq:scaling_relation}
    \begin{array}{c}
    \dot M\propto\mbh^{-1/2}M_*^2R_*^{-1/2},\\
    t(\dot M)\propto \mbh^{1/2}M_*^{-1}R_*^{3/2}.
\end{array}
\end{equation}
For any SMBH mass, the {mass fallback curves} generally look similar in shape, and the flare duration is proportional to the peak time with respect to the disruption, $\tpeak$. 

When it comes to eccentric TDEs, however, the initial orbital energy is non-zero, and one is not able to directly scale the light curves. In particular, this requires us to examine $\tpeak$ and the flare duration separately. 

\subsection{$\tpeak$ in eccentric TDEs}
Using the methods illustrated in Section~\ref{sec:Analyt}, we adopted the $\dmde$\footnote{{the $\dmde$ is extracted at the end of each simulation after running for 100\,$\tdyn\approx1.8\,$d, when the hydrodynamical effects in the tidal tails have ceased to be dominant.}} in STARS library to generate light curves for any eccentricity above $e_{5\%}$. For simplicity, we fix $\beta=1$. An understanding of $\dmde$ allows one to use Kepler's third law to connect the orbital period with the eccentricity of the  orbit: 
\begin{equation}
    \frac{4\pi^2}{P^2}=\frac{G\mbh}{a^3}=\frac{(1-e)^3}{\tdyn^2},
\end{equation}
where $\tdyn$ depends only on the stellar mass and radius and there is a direct relation between $P$ and $e$, which is independent of $\mbh$ for tidal encounters with a fixed $\beta$ value.

   \begin{figure*}
        \centering
        \includegraphics[width=\textwidth]{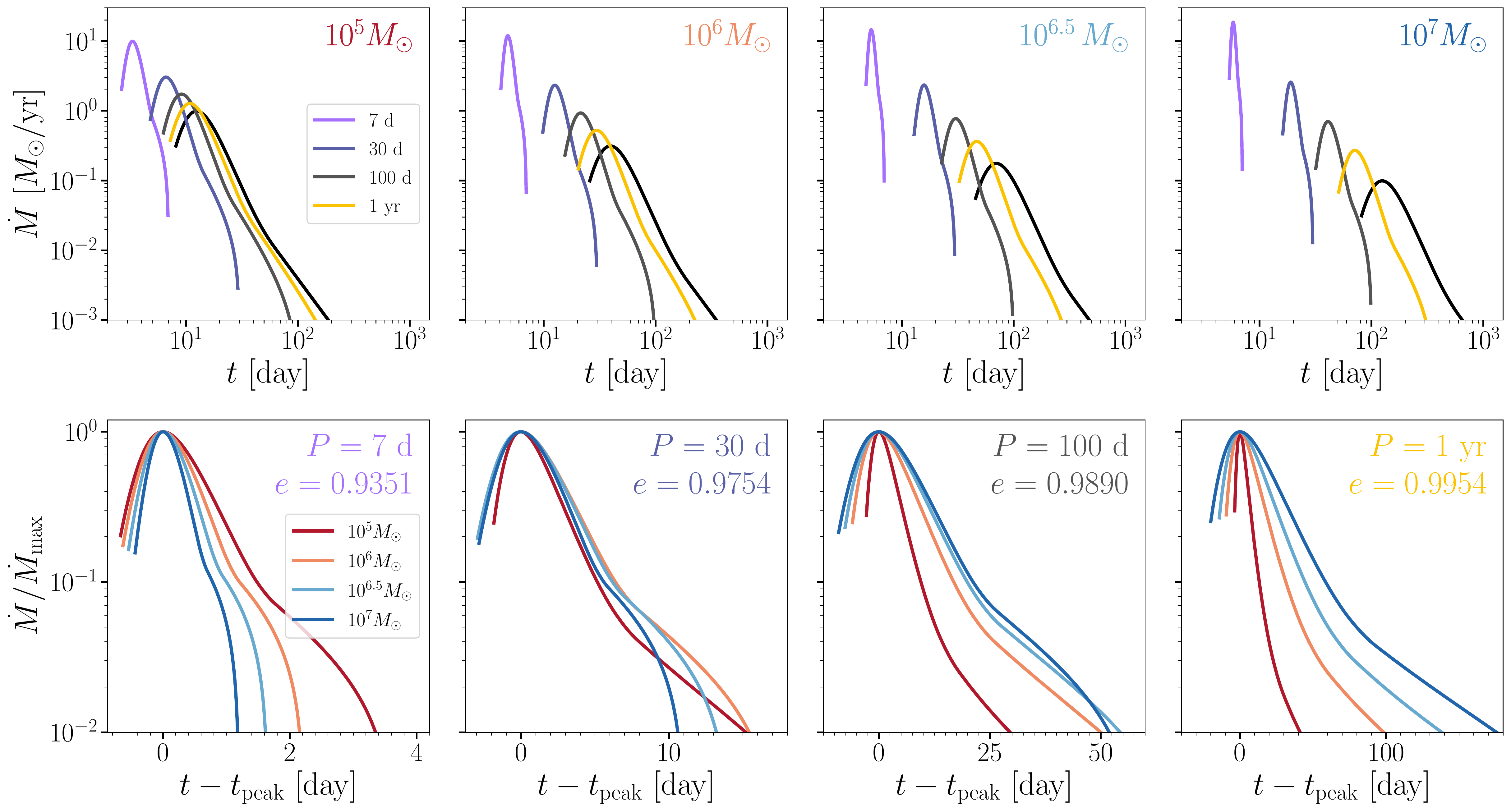}
        \caption{Mass fallback curves from a sun-like star being disrupted by a SMBH  in a variety of eccentric orbits. In all cases $\beta=1$. \textit{Top:} The SMBH mass is fixed in each panel ({$10^5, 10^6, 10^{6.5}$, $10^7\,\mathrm{M_\odot}$}), and the orbital period is varied (7\,d, 30\,d, 100\,d, 1\,yr). Black curves correspond to the widely discussed parabolic encounters. \textit{Bottom:} the orbital period is fixed in each panel, and the SMBH mass is varied. For each curve, the fallback rates are normalized to their maximum value. In order to highlight the differences in the shape, the  $\dot{M}$ curves are aligned them  based on their  $\tpeak$ values.}
        \label{fig:Mdot_T}
    \end{figure*}
    
Figure~\ref{fig:Mdot_T} shows the dependence of the mass fallback rate $\dot M$ with orbital period for a fixed mass of the disrupted SMBH. In the upper panels, each curve corresponds to one orbital period, including the parabolic one (solid black). For $e=e_{5\%}$, the corresponding orbital period is 6.5\,d. All the orbital periods we show in Figure~\ref{fig:Mdot_T} are above this critical value. Each column in the upper panels has a unique central SMBH mass $\mbh$. With a fixed $\mbh$, for lower eccentricity, $\tpeak$ and the duration of a TDE flare both decrease, and therefore the maximum fallback rate, as well as the peak luminosity, increases due to mass conservation, as in all cases we have the same amount of mass stripped from the star.

A rough estimate of $\tpeak$ is given by the fallback timescale of the most bound debris, which lies deepest in the potential of the SMBH. Compared to the CoM of the star, this most bound material has  a relative binding energy of $\sim$$\delta E$, such that
\begin{equation}
\label{eq:tpeak}
\begin{split}
    {\tpeak}&\simeq P(E_\mathrm{CoM}-\delta E)\\
    &\propto\left[q^{-1/3}+\frac{1}{2}\left(\frac{2\pi\tdyn}{P}\right)^{2/3}\right]^{-3/2}.
\end{split}
\end{equation}
Fixing $P$ to be constant, the relation is not self-similar, although $\tpeak$ still monotonically increases as   $q$ is augmented. For the same $q$, $\tpeak$ quickly drops if $P$ decreases, especially for higher mass ratio TDEs (Figure~\ref{fig:Mdot_T}). As such, for eccentric TDEs, we are no longer able to simply rescale the mass fallback rate with $\mbh$.

\subsection{The duration of individual flaring events}
The duration of individual flaring events in repeating  TDEs, {defined as the time that the mass fallback rate is above a given value}, also depends on $\mbh$ and $P$. {In practice, we use the $e$-folding full widths (where ${\md M}/{\md E}$ is greater than $1/e$ of its peak value) of the energy distribution of the tidal debris in terms of the fallback time to characterize the duration of the flare, shown as the shaded regions in Figure~\ref{fig:dMde}. We define the specific binding energy at the inner and the outer edge of the shaded region as $E_-$ and $E_+$. Then when the mass of the SMBH is fixed, varying the orbital period of the TDE will not change $(E_+-E_-)$, i.e., the full width of the energy distribution. The duration $t_\mathrm{w}$ (defined as the difference between the corresponding fallback times for $E_+$ and $E_-$), however, is proportional to $\left(|E_+|^{-3/2}-|E_-|^{-3/2}\right)$.} For long period repeating TDEs, the timescale of the individual flares is well described, as in the case of parabolic encounters, by Equation~(\ref{eq:scaling_relation}).  As such, the flare duration depends on the mass ratio $q$, and goes as $\propto q^{1/2}$. However, when the orbital period is shorter, the fallback curves are significantly altered and the individual flares show a steep decline at times close to the orbital period $P$. This effect is more  pronounced at high $q$ values, as individual flares peak at progressively later times as  $q$ increases. In this regime, the duration of individual flaring events does not increases monotonically with $q$ any longer.
{To explain this complicated dependence on $q$, we first note that both $P(E_+)$ and $P(E_-)$ still increase for more massive SMBHs whose behavior should be similar to that of $\tpeak$, which is characterized by Equation~(\ref{eq:tpeak}). This means that for a sufficiently large $q$ and a sufficiently small $e$, $P(E_+)$ would be asymptotically approaching its upper limit, the orbital period $P$ of the star, and then become insensitive to the change of $q$. Once $P(E_+)$ gets ``saturated'' while $P(E_-)$ keeps increasing for larger $q$, $t_\mathrm{w}$ starts to decrease. This is clearly illustrated in Figure~\ref{fig:dMde}. While the $t_\mathrm{w}$ in parabolic TDEs monotonically increases for more massive SMBHs (black shaded regions), for eccentric TDEs of a short orbital period of 30\,d and when $\mbh>10^6\,\mathrm{M_\odot}$, $P(E_-)$ increases faster than $P(E_+)$ for more massive SMBHs as the latter gets ``saturated'', and $t_\mathrm{w}$ starts to get shorter.}
{Then in the bottom panel of Figure~\ref{fig:Mdot_T}, we compare the duration of the mass fallback for a wider range of models to further illustrate the complicated dependence of the light curves with $q$}. As the period decreases from years to days, progressively less massive black holes start deviating from the characteristic $q^{1/2}$ scaling and, as such, begin to have progressively shorter duration  flares.  In the extreme case of a  very short orbital period (e.g., 7\,d in Figure~\ref{fig:Mdot_T}), the shape of the light curve is altered so significantly  that more massive SMBHs have shorter duration flares. For intermediate values of $P$, the eventual resulting light curves for repeating TDEs are thus expected to depend fairly strongly on the combination of $\mbh$ and $P$. 

There is a positive and a negative side to
this. On the negative side, it implies that one might  be unable to accurately estimate the mass of the SMBH using the properties of the light curve \citep[as done effectively for non-repeating TDEs  by][]{mockler_weighing_2019, Nicholl_2022},
in particular for rapidly repeating flares.  On the positive side,  when we identify a  flare whose duration is shorter than the one predicted by parabolic encounters, which requires an independent estimate of $\mbh$, we  can be fairly certain that the flare is likely to repeat.  This encourages us to present a detailed account of the properties of TDEs that might allow to identify repeating sources  solely on the properties of an individual flare detection.

\section{Phase Space of Repeating Flares}
\label{sec:PhaseSpace}
We have shown that repeating  TDEs in eccentric orbits exhibit shorter and potentially brighter  flares compared to  the widely discussed parabolic encounters. In this section we demonstrate that repeating and single TDEs can be distinguished by using the properties of the  light curve  of a single flare. In particular we show that for $P\lesssim1\,$yr, the expected light curve changes are noticeable enough  in order to effectively predict that a TDE will recur.

For parabolic encounters, the peak timescale can be used to constrain the mass of the disrupting SMBH, given that the luminosity in TDEs closely follows the mass fallback rate \citep{mockler_weighing_2019}. Though $\tpeak$ also depends on $\beta$ and the properties of the disrupted star, these dependencies are not as strong. \citet{law-smith_stellar_2020} show, for example, that for MS stars of varying masses, stellar ages and $\beta$ values disrupted by a  SMBH in a parabolic orbit, $\tpeak$ varies only by  $\lesssim$4. For eccentric TDEs $\tpeak$ should also only weakly depend on the stellar structure of the disrupted star, due to the  minute changes expected  in binding energy distribution throughout the debris when compared to parabolic encounters. 

\begin{figure}
        \centering
        \includegraphics[width=\linewidth]{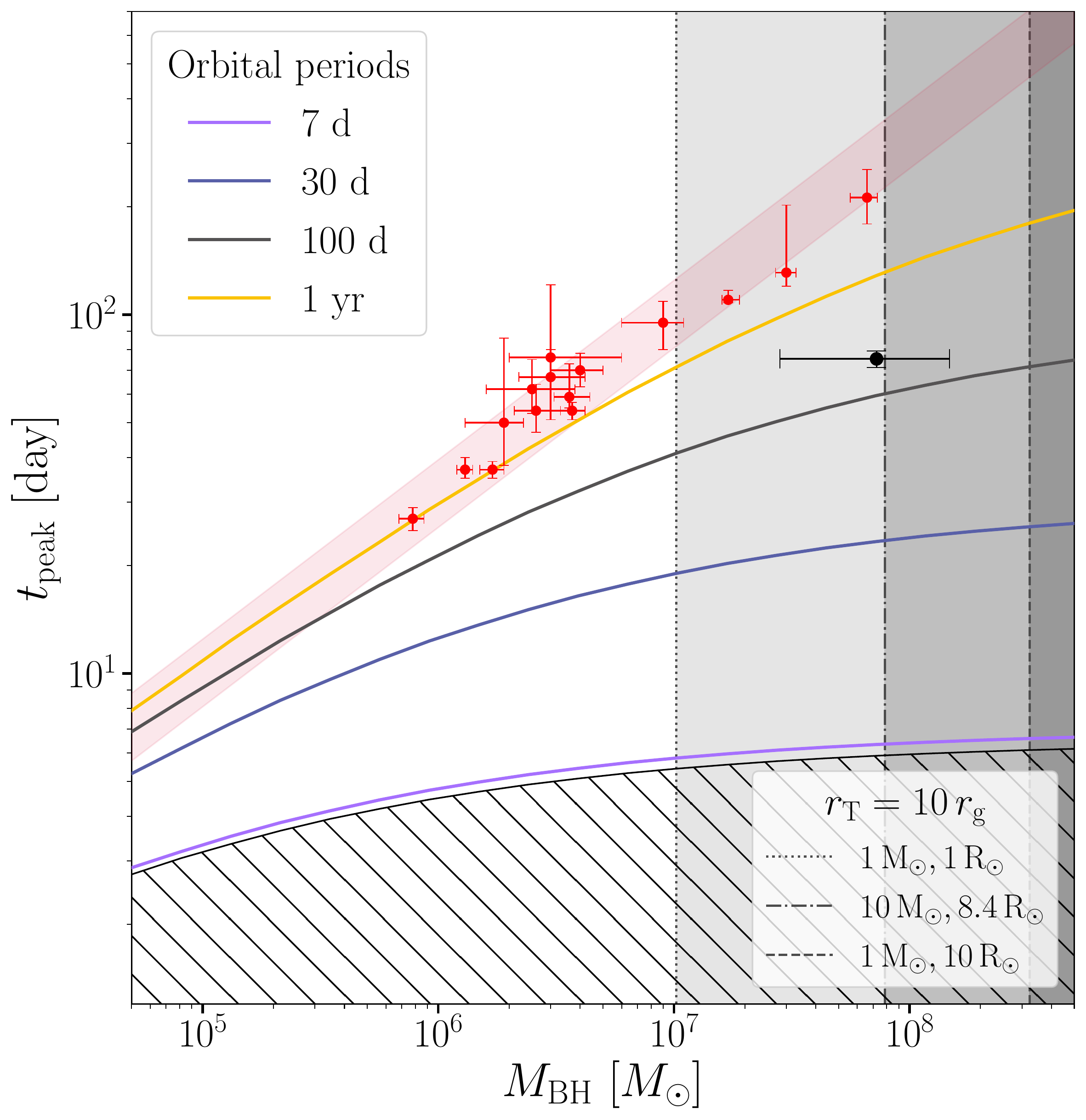}
        \caption{The relation of $\tpeak$ and $\mbh$ for events triggered by the disruption of sun-like stars. The pink shaded region corresponds to parabolic TDEs of varying masses, ages and $\beta$ values. We show the best fit for 14 TDEs from Table 2 in \cite{mockler_weighing_2019} using Modular Open Source Fitter for Transients (\mosfit) as red dots. The colored curves, on the other hand,  show the effects of eccentricity as characterized by the orbital period (7$\,$d, 30$\,$d, 100$\,$d, and 1$\,$yr). We also show the best fit for $\tpeak$ and $\mbh$ for ASASSN-14ko (black dot).  In this case,  we use the estimate for $\mbh$ from \citep{Payne_ASASSN_2021} derived from the galaxy host properties. TDEs in the region {filled with slash lines} have $e\le e_{5\%}$, and the corresponding $\tpeak$ cannot be evaluated using the simulation results of parabolic disruptions. {The dotted, dash-dotted, and dashed vertical lines (from left to right) mark the SMBH masses at which the tidal radius $\rT$ of a sun-like star (1\,$\mathrm{M_\odot}$, 1\,$\mathrm{R_\odot}$), a massive MS star (10\,$\mathrm{M_\odot}$, 8.4\,$\mathrm{R_\odot}$), and an evolved giant star (1\,$\mathrm{M_\odot}$, 10\,$\mathrm{R_\odot}$) is equal to $10\,r_\mathrm{g}$. For more massive SMBHs (in the shaded regions), the relativistic effects cannot be neglected.}} 
        \label{fig:Tp_q}
    \end{figure}

In Figure~\ref{fig:Tp_q}, we show $\tpeak$ as a function of $\mbh$ for the disruption of sun-like stars with  $\beta=1$ in orbits with varying periods. As shown above for stars disrupted in eccentric orbits at a fixed $\mbh$, $\tpeak$ monotonically decreases with decreasing $P$. For reference, the same orbital periods  as in Figure~\ref{fig:Mdot_T} are shown. As highlighted by \citet{mockler_weighing_2019}, $\tpeak$ monotonically rises for parabolic encounters with increasing $\mbh$ and the pink shaded region shows the variation in $\tpeak$ expected for sun-like stars with a wide range of $\beta$ values. This range is calculated from  a rather grazing passage ($\beta=0.5$) to an extremely deep encounter leading to a full disruption ($\beta=4$). 

We plot in Figure~\ref{fig:Tp_q} the $\tpeak$ and $\mbh$ values derived for 14 optical TDEs from Table 2 in \cite{mockler_weighing_2019} as red dots. In these cases,  $\tpeak$ and $\mbh$ were estimated with Modular Open Source Fitter for Transients (\mosfit). The SMBH masses in these events range from $10^6\,\mathrm{M_\odot}$ to $10^8\,\mathrm{M_\odot}$. The data can be found in the Open TDE Catalog \citep{Auchettl_2017, Guillochon_2017} and the detailed selection criteria are described in Section~3.1 of~\cite{mockler_weighing_2019}.  

All these events, as expected, land within the the shaded region in the $\tpeak-\mbh$ phase space of the disruption of a sun-like star, even though the disrupted stars  vary in stellar mass and age \citep{2021ApJ...906..101M,2022ApJ...924...70M}. When $\tpeak$ lies below this shaded region, as in the case of ASASSN-14ko, the  unlucky star is only consistent with having been disrupted by the SMBH in an eccentric orbit and, as such, is predicted  to repeat in the future. It is important to note that in order to predict a repeating event,  the $\mbh$ needs to be independently inferred, using, for example, the host galaxy properties. {If the SMBH causing the TDE is  massive ($\mbh\gtrsim10^7\,\mathrm{M_\odot}$), then the tidal radius of a sun-like star would be close to the event horizon. In this case the mass fallback rates in the STARS library, which are based on non-relativistic simulations, are no longer accurate. In Figure~\ref{fig:Tp_q} we show the critical SMBH masses for a sun-like star (1\,$\mathrm{M_\odot}$, 1\,$\mathrm{R_\odot}$), a massive MS star (10\,$\mathrm{M_\odot}$, 8.4\,$\mathrm{R_\odot}$),\footnote{{This corresponds to the parameters of the most massive star in the STARS library.}} and an evolved star (1\,$\mathrm{M_\odot}$, 10\,$\mathrm{R_\odot}$), defined as where we expect $\rT\leq 10\,r_\mathrm{g}$. Since $\rT\propto \bar\rho^{-1/3}$, lower-mass MS stars undergo strong relativistic effects when disrupted by a SMBH $\gtrsim$$10^7\,$$\mathrm{M_\odot}$. However, the critical SMBH mass for strong relativistic effects can be larger for heavier MS stars and significantly larger for  evolved stars. Although relativistic effects do not significantly alter $\tpeak$  \citep{Stone_2019}, our estimate of $\tpeak$ using the STARS library is not extremely accurate when the encounter is relativistic. In addition,} when $\tpeak$ is present within the {slash-line} region in Figure~\ref{fig:Tp_q}, the corresponding orbital eccentricity is below $e_{5\%}$ and we are thus unable to effectively estimate $\tpeak$ using $\dmde$ extracted from simulations of parabolic TDEs. Having said this, we  expect the vast majority of repeating flares to exist well above the {slash-line region} and be effectively described by using the results of parabolic TDE libraries \citep[e.g.,][]{guillochon_hydrodynamical_2013,law-smith_stellar_2020}.
    
\section{Lessons learned from ASASSN-14ko}\label{Sec:dis}
Here  we discuss the corresponding ramifications of our findings in the context of the recently observed ASASSN-14ko. While repeating nuclear transients have been unveiled, ASASSN-14ko is arguably the most convincing case for being triggered by the disruption of a star in an eccentric orbit. Other events that have been claimed to potentially been associated with repeating TDEs include HLX-1 \citep{lasota_origin_2011, macleod_close_2016, wu_universal_2016, van_der_helm_simulations_2016}, IC 3599 \citep{campana_multiple_2015, Grupe_IC3599_2015}, OGLE16aaa \citep{shu_x-ray_2020}, and eRO-QPE1\&2 \citep{arcodia_x-ray_2021, zhao_quasi-periodic_2022}. 

\subsection{Constraints derived from  light curve fitting}

\begin{figure}
        \centering
        \includegraphics[width=\linewidth]{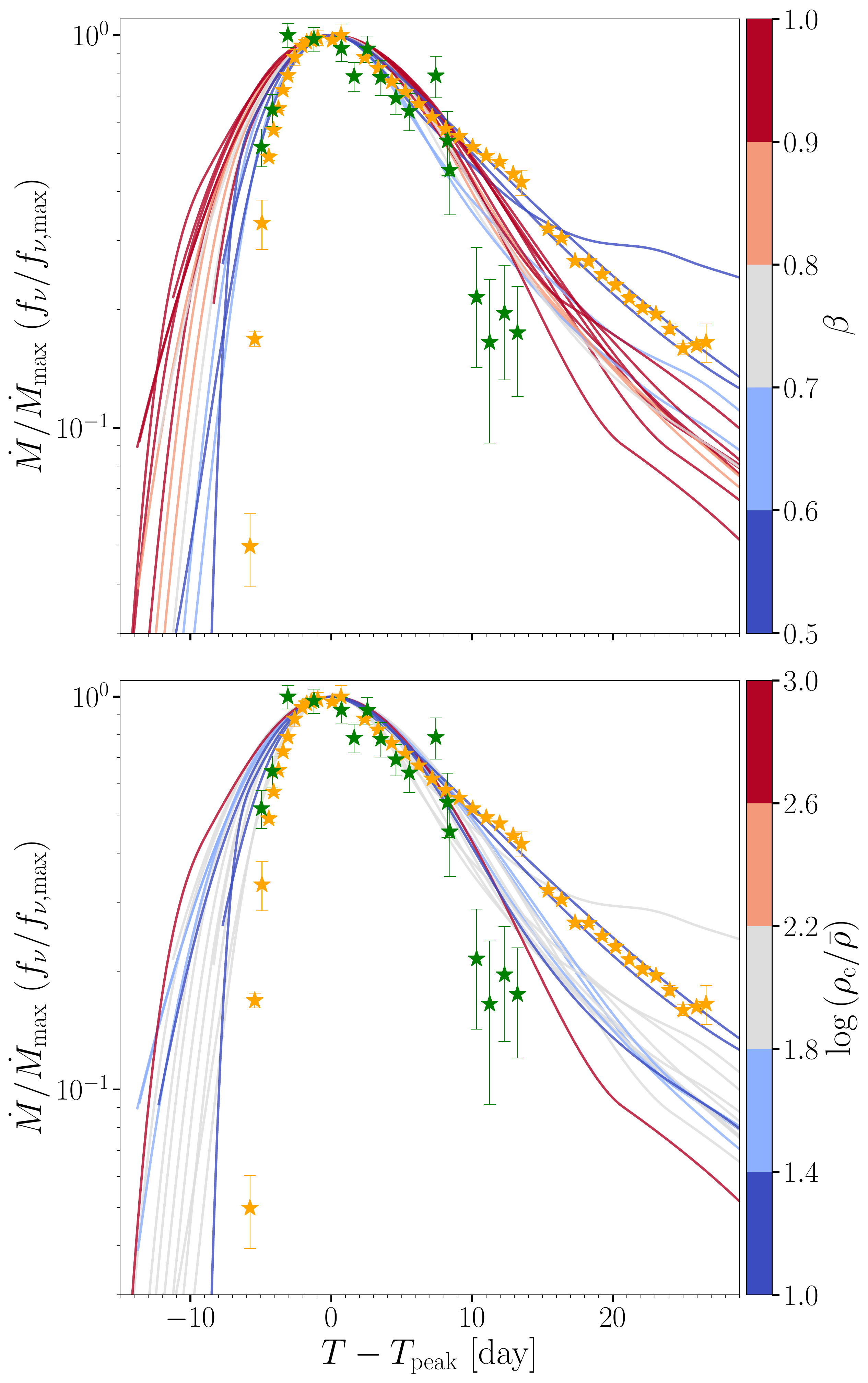}
        \caption{Mass fallback rates for all partial disruption models in the STARS library with $\beta\le1$ and $\Delta M/M \le 0.1$,  where the $\md M/\md E$ distribution has been shifted for an eccentric orbit with $P=114\,$d. These curves are expected to provide a reasonable description of the bolometric luminosity of TDEs \citep{mockler_weighing_2019}. {In the upper (lower) panel, the mass fallback curves are color-coded based on the $\beta$ ($\rho_\mathrm{c}/\bar\rho$) adopted in the corresponding models. As a reference, the normalized multi-band light curves of the November 2018 flare are over-plotted, though they do not strictly follow the bolometric luminosity (thus the mass fallback rate). The orange (green) stars correspond to the light curves in TESS (ASAS-SN $g$).}
       }
        \label{fig:STARS}
    \end{figure}
    
We have shown how the mass fallback curves can be reshaped by putting the disrupted star on eccentric orbits. In this section, we use Modular Open Source Fitter for Transients \citep[\mosfit;][]{guillochon_mosfit_2018} to describe  the multi-band light curves of ASASSN-14ko, with the attempt to constrain the properties of the system. The model takes \texttt{FLASH} simulations of the mass fallback rate of sun-like stars \citep{guillochon_hydrodynamical_2013} as inputs to fit TDE observations.  Details can be found at \cite{mockler_weighing_2019}. The fitting routine was modified for eccentric TDEs and  the orbital period was introduced as a new free parameter, where  the distribution of debris mass $\dmde$ as described in Equation~(\ref{eq:dmde}) was shifted before converting it to the mass fallback rate using Equation~(\ref{eq:mdot}). {We note that the fallback rates in \texttt{MOSFiT} originates from simulations calculated using Newtonian gravity. In Section~\ref{Sec:MF} and \ref{sec:a14ko_structure} we will show that this assumption is valid for ASASSN-14ko.}

\begin{figure*}
        \centering
        \includegraphics[width=\textwidth]{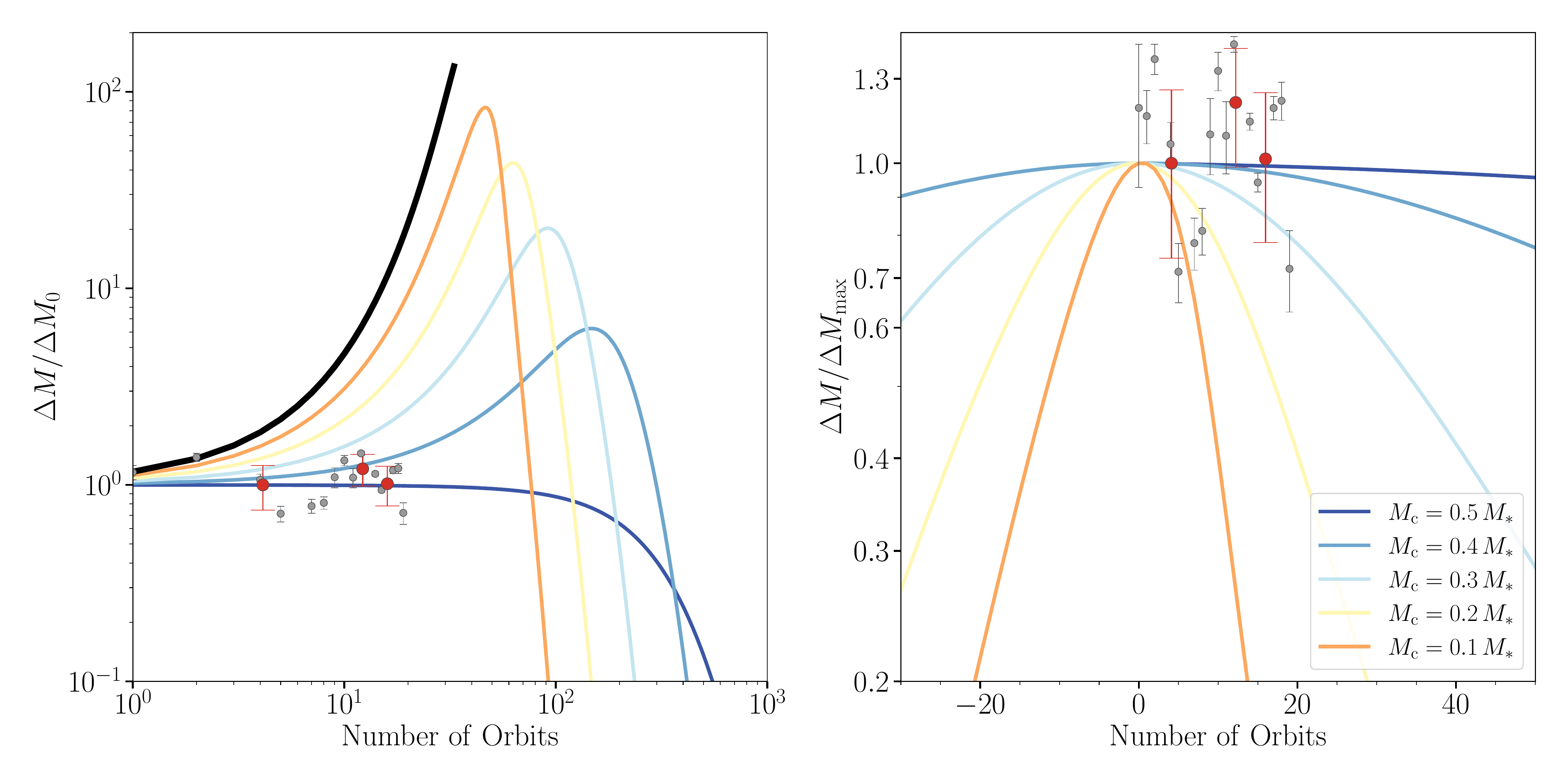}
        \caption{Evolution of mass loss rate for  stars experiencing multiple tidal disruption episodes. The models evolve the stellar structure adiabatically and calculates the  degree of mass loss at each passage based on the results of hydrodynamics simulations. The mass loss in the models are normalized to the first flare (left panel) and the strongest predicted flare (right panel). The curves correspond to the episodic mass transfer from  evolved stars with different core mass fractions and a sun-like star (black curve). All stars have the same mass.  The evolved stars are modeled with nested polytropes while the sun-like star is modeled as a single polytrope (black curve). The polytropic index $\gamma$ of the envelope is set to be $\gamma=5/3$. We over-plot the normalized mass loss rate in each flare of ASASSN-14ko, obtained by fitting the light curves in ASAS-SN $V$-band or ASAS-SN $g$-band as grey dots. The red dots corresponds to the average of the consecutive four flares. To estimate the overall mass loss, for simplicity, we assume black-body emission, no temperature evolution, and a constant energy conversion efficiency $\epsilon$.}
        \label{fig:flare_evo}
\end{figure*}

We select the flare on November 2018 around MJD=58436, which was monitored by Transiting Exoplanet Survey Satellite \citep[TESS;][]{TESS_2015} and All-Sky Automated Survey for Supernovae \citep[ASAS-SN][]{Shappee_2014, Kochanek_2017} at the same time \citep{Payne_ASASSN_2021}. In Figure~\ref{fig:Tp_q}, we show the best fit for $\tpeak$ and $\mbh$ of ASASSN-14ko in the phase space of parabolic and eccentric TDEs (black dot). As argued before,  the comparatively shorter $\tpeak$ of the November 2018 flare of ASASSN-14ko, despite its relatively large $\mbh$, is clearly consistent with this event being a repeating TDE. 

The general structure of the light curve in the ASAS-SN \textit{g} and TESS bands can be effectively described by the {non-relativistic tidal disruption of a star by a SMBH on an eccentric orbit}. Yet, \mosfit~is  unable to effectively model the steep rise of the light curve  with the available library of hydrodynamics simulations. The bluer \textit{g}-band flux also drops faster than in the TESS band, showing significant decrement in the photospheric temperature after the luminosity has reached its peak. This is in contrast with what is observed in other optical TDEs \citep{mockler_weighing_2019}.

Figure~\ref{fig:STARS} shows the mass fallback rates, which are expected to follow the bolometric luminosity, for the most grazing disruptions ($\beta\le1$ and $\Delta M/M \le 0.1$) in the STARS library \citep{law-smith_stellar_2020}, where
the distribution of debris mass $\md M/\md E$ has been shifted for an eccentric orbit with $P=114\,$d. {\citet{law-smith_stellar_2020} found that the rising slope of the mass fallback curve mainly depends on the impact parameter $\beta$ and the ratio of the central and  mean density $\rho_\mathrm c/\bar \rho$, which characterize how centrally concentrated the star is. In the upper/lower panels, the theoretical mass fallback curves are colored-coded based on the two parameters. For these most grazing events in the STARS library, we find that the rising slope has a stronger dependency on $\beta$ than that on $\rho_\mathrm c/\bar \rho$. As has been pointed out in \citet{law-smith_stellar_2020}}, low-$\beta$ encounters show the steepest rises in $\dot{M}$, since only the outermost layers of the star are stripped. In such cases, the spread in binding energy in the outer layers is relatively narrow, which means all the material falls back to pericenter within a similar timescale. {Together with the mass fallback curves, we also plot the normalized flux in TESS and ASAS-SN \textit{g} of the November 2018 flare, with the caveat that the luminosity in broadband optical filters is not equivalent to the bolometric luminosity.\footnote{{The spectral energy distribution (SED) of ASASSN-14ko peaks in ultraviolet \citep[UV;][]{Payne_ASASSN_2021}, such that optical photometry only captures radiation on the Rayleigh-Jeans tail of the SED. As the photosphere cools down, the SED peaks progressively redward, and thus the light curves from optical filters would have a slower rise/decay and peak later than the bolometric luminosity, as well as the mass fallback curve.}} Unfortunately, we do not have high-cadence ultraviolet (UV) observations during the November 2018 flare to constrain the bolometric light curve. ASASSN-14ko shows an unusually fast rise in TESS and ASAS-SN \textit{g}. In the TESS light curve (which has a higher cadence), the brightness increased by a factor of $\sim$20 in only $\sim$6\,days. We expect the rise in the optical light curve to be slower than the rise in the bolometric luminosity because the temperature cools near the peak \citep{Payne_ASASSN_2021}, so we conclude that none of the encounters in the STARS library could reproduce the rising slope of ASASSN-14ko. Nonetheless, extrapolating the mass fallback curves to those from lower-$\beta$ encounters could potentially fix this discrepancy.} Currently, both the library in \mosfit\ and the STARS library have a minimum impact parameter of $\beta=0.5$. The hydrodynamical modeling of more grazing encounters requires higher spatial resolution and, as such, is more computationally expensive. {Another reason that it is difficult to reproduce the rise of ASASSN-14ko is that} the structure of the disrupted star is expected to be modified over multiple passages and is likely to be significantly altered. A more accurate description of the light curve thus requires hydrodynamics simulations of stars experiencing multiple passages, which are difficult to model. {That being said, we expect the leading order changes to the light curve to come from $\beta$, not the stellar structure (see Figure~\ref{fig:STARS}).}

\subsection{Flare evolution over multiple passages}\label{Sec:MF}
After an orbit in which the star experiences a partial disruption, some of the envelope material is removed while some remains marginally bound to the stellar core. The material that is bound is then re-accreted over a few dynamical timescales, which results in the surviving star possessing a hot outer layer envelope.  The process of disruption and the subsequent re-accretion produce a star that has a lower average density and {a shallower density profile}, which makes the star more vulnerable to tidal deformation  on subsequent passages. This regime where stars in an eccentric orbit lose some mass in their first encounter and are only completely destroyed after a large number of passages is largely unexplored with simulations. The reader is referred to \citet{2011ApJ...731..128A} and \citet{guillochon_consequences_2011} for multiple-passage simulations of stars and planets, respectively. There has also been no hydrodynamical modeling of the TDE light curves from multiple-passage encounters, which limits our ability to extract information from the observations. 

The goal of this section is to make use of the simple semi-analytical  model developed by \citet{macleod_spoon-feeding_2013} to gain a deeper physical
interpretation of observations of repeating TDEs.  This formalism, which is formally introduced in Appendix~\ref{app1},  uses an analytical model combining the degree of mass loss from former hydrodynamics simulations of single passages and evolves  the stellar structure adiabatically. At each pericenter passage, the star loses some of its mass and is subsequently restructured under the assumption that it will be completely relaxed after a few dynamical timescales (usually much shorter than the orbital period).  Since the star's orbit is essentially unchanged, we can determine the mass loss at each encounter once we know how the stellar structure evolves as a result of the ensuing mass loss. 

Figure~\ref{fig:flare_evo} shows the predicted evolution of the mass stripping for stars undergoing multiple tidal encounters with a SMBH. The different curves correspond to evolved stars (with different core mass fractions as indicated in the legend) and a sun-like star (black curve). In all models the stars have the same mass. The vulnerability of stars depends sensitively on the  core mass fraction, with stars without cores being the most vulnerable and being destroyed rapidly. In all models with a initial core mass fraction less than 0.5, as the star loses mass to the SMBH, its radius increases and it becomes increasingly more vulnerable to tidal disruption. After a few passages, usually when the core mass is comparable to the envelope mass, the star's radius shrinks as the envelope is depleted. 
A comparison with ASASSN-14ko shows that the amount of mass estimated to have been accreted from passage to passage does not markedly increase or decrease over the past $\sim$20 flares, insinuating at the disruption of an evolved star with a high core mass fraction. It is to this issue that we now turn our attention. 

\subsection{Constraints on the structure of the disrupted star}\label{sec:a14ko_structure}
In the first 12 flares since ASASSN-14ko was discovered and identified, ASAS-SN monitored 8 of the light curves in $V$-band. From the 13-th flare on, ASAS-SN started monitoring  them in $g$-band. Making use of this data we deduce the accreting mass loss $\Delta M$ in each flare, which we assume is directly related to the mass loss of the star. The normalized $\Delta M$ for each flare is over-plotted in Figure~\ref{fig:flare_evo}. To estimate the bolometric luminosity, we fit the light curve in either the ASAS-SN $V$-band or $g$-band for each flare in order to derive the corresponding single band luminosity. We assume black-body radiation and a constant emitting photosphere temperature throughout each flare, $T=10^{4.5}\,$K, which was derived from the multi-band observation of the May 2020 flare \citep{Payne_ASASSN_2021}. The radiation efficiency is also assumed to remain constant $\epsilon=0.01$ \citep{Dai_2018, Jiang_2016}. {But since we focus on the trend of the evolution over multiple flares in this section, the value of this constant will have no impact on our results.}
In the left panel of Figure~\ref{fig:flare_evo}, the analytical flare evolution models are normalized to the mass loss estimated from the first flare. The observed $\Delta M$ (normalized by the estimated $\Delta M$ in the first observed flare to get rid of the impact due to the uncertainty in $\epsilon$) clearly oscillates, while the average $\Delta M$ over every six consecutive flares shows no significant variation. If ASASSN-14ko emerged near the time it was first detected, the lack of evolution of $\Delta M$ from flare to flare would rule out the disruption of an evolved star  with a core mass fraction $\lesssim$0.3. This is because these models predict  an increase in $\Delta M$ by a factor of $\gtrsim$2 over the first 20 flares, which is clearly not observed.  Because ASAS-SN was not able to constrain the emergence of the event \citep{Payne_ASASSN_2021, Payne2022}, we also show the results for the most conservative model, in which  the evolution of the mass loss in all the models is the slowest. This corresponds to observing the event near peak (right panel of Figure~\ref{fig:flare_evo}). While a higher core mass fraction is still preferred by the data, for a star with core mass fraction of $0.3$, $\Delta M$ is only reduced by 30\% with respect to its peak, which is in marginal agreement with the observations. Models with  core mass fraction $<$0.3 are still not consistent with the observations. {Beyond our simple adiabatic model, hydrodynamics (see Section~\ref{subsec:multi_pass}) and potential relativistic effects \citep[e.g.,][]{Gafton_2015} only increase the fraction of the mass loss $\Delta M/M$ in each encounter} and thus accelerate the evolution of the disruption. In conclusion, an evolved star with a core mass fraction  $\gtrsim$0.3  is likely neededto reproduce the flare evolution of ASASSN-14ko.

{If ASASSN-14ko originates from an evolved star, it will not experience strong relativistic effects. Near a SMBH of $\sim $$7\times10^7\,$$\mathrm{M_\odot}$, the tidal radius $\rT$ of a sun-like star is $\sim$2.8\,$r_\mathrm{g}$, indicating a highly relativistic encounter even for grazing interactions($\beta\lesssim0.5$). 
But for an extended star, the tidal radius can be much greater. \citet{macleod_tidal_2012} used MESA to simulate a solar-metallicity star of 1.4\,$\mathrm{M_\odot}$ and found that the core mass fraction will not reach 0.3 until the star evolves and reaches the tip of the red giant (RG) branch. The core mass fraction then keeps increasing following the star's evolution track through the horizontal branch (HB) and the asymptotic giant branch (AGB), and for most of the time when the core fraction is $\gtrsim$0.3, the star stays in the HB and has a typical radius of $\sim$10\,$\mathrm{R_\odot}$. The characteristic tidal radius is then a factor of $\sim$10 larger than that of a sun-like star, or more specifically, $\rT\approx 28\,r_\mathrm{g}$. And for grazing encounters with small $\beta$, as expected for ASASSN-14ko ($\beta\lesssim0.5$), the actual $\rp$ would be $\gtrsim$50\,$r_\mathrm{g}$. We thus argue that for ASASSN-14ko, the relativistic effects are not significant.}

{AGB stars have even higher core mass fraction ($\sim$0.5) and more extended envelopes ($\sim$$10^2$$\,\mathrm{R_\odot}$). But such a large stellar radius would be problematic, because $\rT$ would be comparable with the semi-major axis of a 114-day-orbit ($\rT\approx a\approx 3\times10^{15}\,$cm), which means the orbit is nearly circular. In this case, the mass transfer would resemble Roche lobe overflows in stellar binaries, and we would expect continuous emission instead of periodic flares.}

\subsection{Constraints on the stellar mass}
An estimate for the total mass of the disrupted star can be obtained by summing over all of the $\Delta M$ from past flares, although the outcome will depends sensitively on  radiation efficiency. Adopting a typical value of $\epsilon=0.01$, the mass fallback to the SMBH is about 0.04 to 0.08\,$\mathrm{M_\odot}$ for each flare, which is  consistent with that expected for grazing encounters of evolved stars with  tenuous envelopes. The total mass loss deduced from all observed individual flares is thus $\lesssim$1\,$\mathrm{M_\odot}$, which provides us with a rough lower limit on the stellar mass. 

Such a low mass loss rate may draw concerns that there might not be sufficient  material surrounding the SMBH in order to effectively reprocess the energetic photons from the accretion disk and generate the observed optical emission. To estimate the optical depth, we adopt the accretion-disk-powered wind model presented in \citet{Roth_Xray_2016} \citep[and similar to the calculations in][]{Dai_2018}, 
which provides an estimate for the  opacity needed to effectively reprocess the high-energy radiation. {This reprocessing model also qualitatively applies for flares whose emission is powered by stream-collision \citep[e.g.,][]{Lu_2020, Bonnerot_2021}.} The density profile in such a wind-driven outflow  is assumed to be proportional to $1/r^{2}$,\footnote{{If the reprocessing layer is fully supported by radiation pressure, the density profile would be proportional to $1/r^3$ \citep[e.g.,][]{Loeb_1997}, leading to a slightly higher opacity.}} such that the optical depth through the wind can be written as 
\begin{equation}
    \label{eq:tau}
    \tau_T=\frac{\kappa_T M_\mathrm{env}}{4\pi r_\mathrm{in}r_\mathrm{out}}.
\end{equation}
where $\kappa_T$ is the Thomson opacity for electron scattering, $M_\mathrm{env}\approx \Delta M$ is the mass of the outflow and $r_\mathrm{in}$ and $r_\mathrm{out}$ are the inner and outer radius of the wind. {\citet{Payne_2022b} estimated the blackbody radius of the photosphere during a single flare using UV data, which expands from $10^{14.2}$ to $10^{14.9}$\,cm as the luminosity rises to its maximum. This means $r_\mathrm{in}$ of the reprocessing layer where the wind originates is $\lesssim$10$^{14.2}$\,cm and $r_\mathrm{out}\approx10^{14.9}$\,cm when the UV luminosity reaches its peak value. This yields $\tau_T \approx20 (\Delta M/{0.05\,\mathrm{M_\odot}})$, which is sufficient to reprocess a significant fraction but not all of the radiation emanating from the hotter disk \citep[$\tau_T\gtrsim20$;][]{Roth_Xray_2016}. This is consistent with the detection of the accompanying  X-ray emission in this event \citep{Payne2022}. While we do not seek to model the radiative transfer process in detail, this simple estimate shows that such a low stripped mass per encounter is broadly consistent with the observations and implies that our assumption for the radiation efficiency $\epsilon=0.01$ is also reasonable. If $\epsilon$ is significantly higher, we would expect a much smaller $\Delta M$, which will be  insufficient to reprocess the high-energy radiation.}

\subsection{Constraints on the mean stellar density}\label{sec:a14ko_dens}
The central concentration and average density of the disrupted star have a strong impact on the duration and shape of TDE light curves \citep{law-smith_stellar_2020}. Similarly, in the repeating TDE case, the duty cycle (defined as the ratio of the flare duration and the recurrence period) depends sensitively on the stellar density. According to the hydrodynamics simulations presented in \citet{law-smith_stellar_2020}, given the SMBH mass, the mass fallback timescale depends mainly on the impact parameter $\beta$ and on the central concentration of the star $\rho_\mathrm c/\bar\rho$. Practically, we can approximate the duration of the flare as the timescale between the peak and when the least bound material returns to the pericenter (given that the rise timescale is very short). And since the least bound material returns to the pericenter at about an orbital period, $P$, the duration for each flare can be approximated as  $P-\tpeak$. The duty cycle $D$ is roughly given by 
\begin{align}\label{eq:duty_cycle}
    D&\simeq1-\frac{\tpeak}{P}\nonumber\\
    &= 1-\left[1+\left(\frac{t_{\mathrm{peak},0}}{P}\right)^{-2/3}\right]^{-3/2},
\end{align}
where $t_{\mathrm{peak},0}$ is the corresponding peak timescale for a star in a parabolic orbit with the same pericenter distance. Since, on average, each flare lasted for $\gtrsim$55\,d \citep{Payne_ASASSN_2021}, $D$ has a lower limit of $\sim$0.48. {\citet{law-smith_stellar_2020} provided an analytical formula for $\tpeak$ at any $\beta\gtrsim0.5$ and $\rho_\mathrm c/\bar\rho\lesssim10^3$. Adopting $\beta=0.5$, we find the estimated duty cycle using Equation~(\ref{eq:duty_cycle}) increases for higher $\rho_\mathrm c/\bar\rho$ because the mass fallback rates of more centrally concentrated stars tend to peak earlier. Increasing $\rho_\mathrm c/\bar\rho$ from the solar value ($\sim$$10^2$) to the maximum in the STARS library ($\sim$$10^3$), $D$ increases from 0.32 to 0.38, still fairly far from the observed quantity. Since \citet{law-smith_stellar_2020} did not perform simulations for evolved, giant stars, their empirical fitting formula might be inaccurate for $\rho_\mathrm c/\bar\rho\gtrsim10^3$.} Nevertheless, this unusually high duty cycle again indicates an evolved star whose envelope has bloated by a factor of few when compared to a mildly evolved sun-like star ($\rho_\mathrm c/\bar\rho\approx 10^{2}$). 

The possible delay due to circularization also influences the flare \citep[e.g.][]{2014ApJ...783...23G}. The stripped material bound to the SMBH falls back to  pericenter at a rate given by Equation~(\ref{eq:mdot}) and Equation~(\ref{eq:dmde}). The returning gas does not produce a flare immediately, but needs to be circularized and form an accretion disk before falling into the SMBH \citep[e.g.,][]{Ramirez-Ruiz_2009, Bonnerot_2017, Bonnerot_2021, Andalman_2022}. Once the gas enters a quasi-circular orbit, which is drastically aided by general relativistic effects \citep{Guillochon_dark_2015, Bonnerot_2021}, it will be accreted within a viscous timescale. The viscous timescale is usually short compared to the typical orbital timescale of the bulk of the stellar debris \citep{rees_tidal_1988,macleod_tidal_2012}.
However, when the {delay due to circularization} is comparable to the fallback timescale, the resulting TDE light curve will be  drastically flattened  \citep{mockler_weighing_2019}, resulting in a much shallower rise in luminosity {than the one observed in ASASSN-14ko. In addition, the bright, fast evolving X-ray emission in ASASSN-14ko \citep{Payne2022, Payne_2022b} also indicates efficient disk formation.} Hence for ASASSN-14ko, the {circularization delay and the viscous timescale are both} expected to be negligible. {Since the pericenter distance $\rp$ is expected to be $\gtrsim$50\,$r_\mathrm{g}$, the relativistic precession is likely not effective in driving the circularization process. Dissipation with disrupted material from previous passages accumulated near the pericenter might accelerate disk formation.} The ratio of the viscous timescale to the orbital period of the progenitor ($P=114\,$d) is given by \citet{macleod_tidal_2012} 
\begin{equation}
\label{eq:tvis}
\frac{t_\mathrm{vis}}{P} \simeq 10^{-2}\left(\frac{\beta}{0.5}\right)^{-3/2}\left(\frac{\alpha}{0.1}\right)^{-1}\left(\frac{\bar\rho_*}{\bar\rho_\odot}\right)^{-1/2},
\end{equation}
where we have adopted a standard $\alpha$-disk  model \citep{Shakura_1973} and assume a geometrically thick disk $(H/R)^2\approx0.1$ as expected for a TDE \citep[e.g.,][]{Dai_2018}.
For an evolved star with only one percent of the solar density, the viscous delay should still be relatively short, provided circularization occurs promptly. An even lower mean density might be difficult to reconcile with the observations, and a lower limit of the rise timescale would then be set by the viscous delay rather than the mass fallback timescale.

\section{Synopsis and Conclusions}

\subsection{Summary}
\begin{itemize}
 \item A star in an eccentric orbit around a SMBH can be tidally stripped at each pericenter passage, leading to a repeating TDE. Making use of  both analytical  and hydrodynamical methods, we show that the overall tidal deformation of a star is similar for both eccentric and parabolic orbits, when the orbital eccentricity is above a particular  critical value (Figure~\ref{fig:analy_cri_ecc}). This allows one to directly calculate the mass fallback rate using existing simulation results for parabolic TDEs (Figure~\ref{fig:Mdot_T}).

\item The structure of flares in repeating TDEs are studied here using  models for the disruption of sun-like stars from the Stellar TDEs with Abundances and Realistic Structures (STARS) library \citep{law-smith_stellar_2020}. Unlike parabolic TDEs, whose peaking timescale with respect to the disruption is  $\propto \mbh^{1/2}$, eccentric TDEs  show more complicated  light curves, with $\tpeak$  depending on both on $\mbh$ and $P$. Still, $\tpeak$ monotonically increases with  $q$, but cannot exceed the orbital period of the star. For stars orbiting  SMBH in  relatively short orbital periods, $\tpeak$  will approach $P$, and the mass fallback will be significantly squeezed. As such, we expect rather short duration flares for a given SMBH, whose light curves will be clearly distinguishable from parabolic TDEs. The presence of such short-lived flares  can then be used to predict repeating sources for SMBH whose mass can be independently measured (Figure~\ref{fig:Tp_q}).  
  
\item The evolution of the structure of a star experiencing  multiple disruptions is studied by assuming  mass loss takes place adiabatically, which we argue provides a robust upper limit to the survivability of the star. Using this formalism we conclude that only evolved stars are able to produce more than tens of flares before full disruption. The  survivability of stars depends crucially on the core to envelope mass ratio, with stars with tenuous envelopes being able to persist for up to hundreds of passages (Figure~\ref{fig:flare_evo}).

\item The repeating TDE candidate ASASSN-14ko \citep{Payne_ASASSN_2021} is discussed within the theoretical framework outlined in this paper. ASASSN-14ko shows recurrent flares every $\sim$114\,d since its discovery in November 2014. Our mass fallback rate libraries adapted to eccentric orbits (\texttt{MOSFiT} and STARS library) are able to qualitatively describe the bolometric luminosity of the best sampled flare (Figure~\ref{fig:STARS}). Yet, these unperturbed models are not able to effectively capture the steep rise of the best sampled flare. This is not surprising as we expect the outer layers of the stars to be drastically altered from previous encounters and be highly vulnerable to grazing encounters. Since stars with lower $\beta$ show steeper rises, we expect rapidly raising light flares from these inflated stars on fairly grazing orbits.  In addition, the survival after over twenty flares, the observed mild flare luminosity evolution (Figure~\ref{fig:flare_evo}), and the relatively long duty cycle all favor a moderately massive ($M\gtrsim 1\,\mathrm{M_\odot}$), extended ({most likely $\sim$10\,$\mathrm{R_\odot}$}), evolved star as the progenitor of this repeating TDE. 
\end{itemize}

\subsection{The need for hydrodynamics simulations of multiple tidal interactions}
\label{subsec:multi_pass}
While we can use a periodic mass stripping model of an evolved star to qualitatively describe both the single flare behavior and the long-term evolution of ASASSN-14ko, we are not yet able to reproduce all observational aspects quantitatively. This fact suggests the need for conducting hydrodynamics simulations of multiple tidal interactions and understanding the evolution of the stellar structure over multiple passages.

In this paper we calculate the mass fallback rate for unperturbed stars using  hydrodynamics simulations of single eccentric passages, and trace the evolution of stellar structure assuming mass stripping occurs adiabatically. As the star experiences mass stripping, it re-accretes the majority of the stellar material. During this re-accretion, the free-falling material builds a standing accretion shock when it encounters the star's unperturbed surface, and  kinetic energy is effectively converted into thermal energy \citep{guillochon_consequences_2011}. 
A hot outer layer is formed in the star's envelope, which extends significantly beyond the initial stellar radius \citep{macleod_spoon-feeding_2013}. On the other hand, the marginally bound material also transfers significant amount of angular momentum towards the star when it is re-accreted. This leads to a rapid spin-up of the star's outer layers \citep{guillochon_consequences_2011}. 

The influence of shock heating  produces an extended atmosphere with a low mean density, which is then more vulnerable to tidal disruption in subsequent encounters. Simple analytical adiabatic models, like the ones used  here, cannot effectively  reproduce the build-up of such a diffuse layer. While $\tpeak$ should not vary too significantly from passage to passage, we expect  the light curve to differ from the ones calculated using undisturbed stellar  models, such that the flare luminosity should evolve more rapidly as was predicted by our analytical models. In addition, the disrupted star should survive fewer orbits when the tidal dissipation is self-consistently  included. Radiative cooling is also expected to be important in the most tenuous outer layers of the hot atmosphere before the subsequent strong tidal encounter and should not be  neglected in simulations. These effects on stars experiencing multiple disruptions  remain an open question and they should be included in   future simulations. Unfortunately, modeling the disruption of stars in multiple orbits hydrodynamically is not currently computationally affordable, because the corresponding orbital periods are too large when compared to the dynamical timescale of stars. For ASASSN-14ko, $P\approx 3 \times 10^3\,\tdyn$. \\

Our understanding of TDEs has come a long way since their prediction  more than thirty years ago \citep{rees_tidal_1988}, but these
nuclear sources continue to offer major puzzles and challenges. Repeating TDEs, such as  ASASSN-14ko, provide us with an exciting
opportunity to study new regimes of tidal interactions. Space- and ground-based observatories over the coming years should allow us to uncover the detailed physics of these most remarkable sources.\\

{We thank the anonymous referee for a thoughtful and detailed report.} We have substantially benefited from the insight and perspective of K. Auchettl and M. MacLeod. B.M. is grateful for the AAUW American Fellowship and the UCSC Presidents Dissertation Fellowship. The  group at UCSC is grateful for support from the Heising-Simons Foundation, NSF (AST-1615881, AST-1911206 and AST-1852393), Swift (80NSSC21K1409, 80NSSC19K1391) and Chandra (GO9-20122X). The UCLA group acknowledges the partial support from NASA ATP 80NSSC20K0505 and thanks Howard and Astrid Preston for their generous support. S.R. thanks the Nina Byers Fellowship, the Charles E. Young Fellowship,  the Michael A. Jura Memorial Graduate Award for support and the Thacher Summer Research Fellowship. D.M. acknowledges partial support from an NSF graduate fellowship DGE- 2034835, and the Eugene Cota-Robles Fellowship.
   
We acknowledge use of the {\it lux} supercomputer at UCSC (AST 1828315), and the HPC facility at the University of Copenhagen (VILLUM FONDEN 16599).\\

\software{\texttt{astropy} \citep{Astropy_2013, Astropy_2018}, \texttt{FLASH} \citep{fryxell_flash_2000}, \texttt{matplotlib} \citep{Matplotlib_2007}, \texttt{MESA} \citep{Paxton_mesa_2011}, \texttt{MOSFiT} \citep{guillochon_mosfit_2018}, \texttt{scipy} \citep{Scipy_2020}, \texttt{yt} \citep{yt_2011}.}

\appendix
\section{Stellar Disruption over Multiple Passages}\label{app1}

Following \citet{macleod_spoon-feeding_2013}, we estimate the overall mass stripped by the SMBH, $\Delta M$, after a given  encounter by adopting the approximating formula derived from simulation results \citep{macleod_tidal_2012, guillochon_hydrodynamical_2013},
\begin{equation}
\label{eq:nested_poly}
    \Delta M(\beta)=f(\beta)\left(\frac{M_*-M_\mathrm{c}}{M_*}\right)^2M_*,
\end{equation}
where $M_\mathrm c$ is the core mass and
\begin{equation}
    f(\beta)=\left\{
    \begin{array}{ll}
        0 & \mathrm{if}\ \beta < 0.5,\\
        \beta/2-1/4 & \mathrm{if}\ 0.5\le\beta\le2.5,\\
        1 & \mathrm{if}\ \beta > 2.5.
    \end{array}
    \right.
\end{equation}
{In the first encounter, we set $\beta=0.501$.} This formalism can be applied to  both main sequence star and giants. Main sequence stars are modeled as single polytropes. Evolved stars have a condensed core of various size and a cool, convective envelope, so we treated them as nested polytropes with initial core mass fractions $M_\mathrm{c}/M_{\ast}$ ranging from $0.1-0.5$. For nested polytropes, the mass-radius relation can be written as $R_*\propto M_*^{\xi_\mathrm{ad}}$, where $\xi_\mathrm{ad}$ is given by the approximate formula \citep{Hjellming1987}
\begin{equation}
    \xi_\mathrm{ad}\approx\frac{1}{3-n}\left(1-n+\frac{M_\mathrm{c}}{M_*-M_\mathrm{c}}\right).
\end{equation}
The polytropic index was taken to be $\gamma=5/3$ ($n=1.5$), which corresponds to a fully convective atmosphere. 

A disrupted star in an eccentric orbit loses mass on a timescale faster than the thermal timescale but slower than the dynamical timescale of the star. In these cases, the structure of the star will evolve adiabatically as assumed by the model outlined above. The validity of this assumption was studied  by  
\citet{macleod_spoon-feeding_2013}. These authors  calculated the changes to stellar properties as the star loses mass  using the \texttt{MESA} stellar evolution code, where they allowed the star to adjust to the mass loss continuously. Their simulation results showed that only for highly extended giant star models the  outer layers of the star would evolve faster than what nested polytropes would predict. These models, however, are still incomplete as they don't take into account the energy injection from tides, which are shown to be important in altering the structure of the object. This was clearly illustrated by \citet{guillochon_consequences_2011} in the context of planet disruption. These authors performed hydrodynamics simulations of a tidally stripping giant planet ($n=1$) tidally stripping around a sun-like star. For a similar initial impact parameter $\beta\approx 0.5$, the authors showed that the planet would be completely destroyed in roughly a few  orbits, during which the mass loss rate increases by orders of magnitude at a rate that is much larger than the one  predicted by  our simple adiabatic polytrope model. In this paper we remain mindful of the limitations of these models.

\bibliography{ref.bib}
\end{document}